\documentclass[12pt]{article}
\usepackage{jheppub}
\usepackage{mathtools,amssymb,amsthm}
\usepackage[utf8]{inputenc}
\usepackage{enumitem}
\usepackage{xcolor}
\usepackage{graphicx}
\usepackage{multirow}
\usepackage[normalem]{ulem}
\usepackage{longtable}
\allowdisplaybreaks
\usepackage{slashed}
\usepackage{bbm}

\hypersetup{
    colorlinks = true,
    allcolors = {olive}
}

\newcommand{\be}{\begin{equation}}
\newcommand{\ee}{\end{equation}}
\def \bea {\begin{eqnarray}}
\def \eea {\end{eqnarray}}
\def\bal#1\ea{\begin{align}#1\end{align}}
\def\bad#1\ead{\begin{aligned}#1\end{aligned}}
\def\bg#1\eg{\begin{gather}#1\end{gather}}
\def\bm#1\em{\begin{multline}#1\end{multline}}
\def\bmd#1\emd{\begin{multlined}#1\end{multlined}}

\newcommand{\ignore}[1]{}

\def\d{\delta}

\def\l{\lambda}
\def\L{\Lambda}
\def\m{\mu}
\def\n{\nu}

\def\r{\rho}
\def\s{\sigma}

\def\X{\Xi}

\def\i{\mathrm{i}}

\def \bal#1\eal  {\begin{align} #1 \end{align}}
\def \bga#1\ega  {\begin{gather} #1 \end{gather}}
\def\({\left(}
\def\){\right)}
\def\[{\left[}
\def\]{\right]}
\def\<{\left\langle}
\def\>{\right\rangle}
\def\d{\mathrm{d}}

\newcommand{\eim}{\end{itemize}}
\newcommand{\beq} {\begin{equation}}
\newcommand{\eeq} {\end{equation}}
\newcommand{\bc}{\begin{center}}
\newcommand{\ec}{\end{center}}

\newcommand{\nn} {\nonumber\\}

\newcommand{\pd} {\partial}

\newcommand{\ai}{{\alpha}}
\newcommand{\bi}{{\beta}}
\newcommand{\gi}{{\gamma}}
\newcommand{\di}{{\delta}}
\newcommand{\ri}{{\rho}}
\newcommand{\si}{{\sigma}}
\newcommand{\li}{{\lambda}}

\newcommand{\ei}{{\eta}}

\newcommand{\oi}{\omega}

\newcommand{\Gi}{\Gamma}

\newcommand{\Li}{\Lambda}

\newcommand{\bv}{{\bar{v}}}
\newcommand{\br}{{\bar{r}}}
\newcommand{\bx}{{\bar{x}}}

\newcommand{\mE}{\mathcal{E}}
\newcommand{\mM}{\mathcal{M}}
\newcommand{\mX}{\mathcal{X}}
\newcommand{\pr}{{\prime}}
\newcommand{\prpr}{{\prime\prime}}

\begin{document}

\title{Pole skipping in holographic theories with gauge and fermionic fields}
\author[a]{Sirui Ning,}
\author[b,c]{Diandian Wang,}
\author[b]{and Zi-Yue Wang}
\affiliation[a]{{The Rudolf Peierls Centre for Theoretical Physics, University of Oxford,\\Oxford OX1 3PU, UK}}
\affiliation[b]{Department of Physics, University of California,
Santa Barbara, \\CA 93106, USA}
\affiliation[c]{Center for the Fundamental Laws of Nature, Harvard University,
      Cambridge,\\ MA 02138, USA}
\emailAdd{sirui.ning@physics.ox.ac.uk}
\emailAdd{diandianwang@fas.harvard.edu}
\emailAdd{zi-yue@physics.ucsb.edu}

\abstract{
Using covariant expansions, recent work showed that pole skipping happens in general holographic theories with bosonic fields at frequencies $\mathrm{i}(l_b-s) 2\pi T$, where $l_b$ is the highest integer spin in the theory and $s$ takes all positive integer values. We revisit this formalism in theories with gauge symmetry and upgrade the pole-skipping condition so that it works without having to remove the gauge redundancy. We also extend the formalism by incorporating fermions with general spins and interactions and show that their presence generally leads to a separate tower of pole-skipping points at frequencies $\mathrm{i}(l_f-s)2\pi T$, $l_f$ being the highest half-integer spin in the theory and $s$ again taking all positive integer values. We also demonstrate the practical value of this formalism using a selection of examples with spins $0,\frac{1}{2},1,\frac{3}{2},2$.
}

\maketitle


\section{Introduction}



Pole skipping refers to the multivalued nature of Green's functions at special points in the momentum space where lines of zeros intersect lines of poles \cite{Grozdanov:2017ajz,Blake:2017ris,Blake:2018leo}. In holographic theories at large $N$ and finite temperature $T$, this phenomenon can be understood through the existence of extra ingoing modes at the black hole horizon, resulting in non-uniqueness of the bulk solution that determines the holographic retarded Green's function \cite{Blake:2018leo}. The first example of pole skipping was found in Einstein gravity, where it happens at frequency $\omega=\i \l_L$ and momentum $k=\i\l_L/v_B$ \cite{Grozdanov:2017ajz,Blake:2018leo}, $\l_L$ being the Lyapunov exponent \cite{Shenker:2013pqa,Shenker:2013yza,Shenker:2014cwa,Maldacena:2015waa} and $v_B$ being the butterfly velocity \cite{Roberts:2014isa,Roberts:2016wdl} for this theory. 
One can therefore say that the analytic structure of the Green's function contains some information about the chaotic properties of the quantum system, especially true when the stress tensor dominates chaos \cite{Mezei:2019dfv,Choi:2020tdj,Blake:2021wqj}.

For bulk theories containing bosonic fields, it was noticed through numerous examples that pole skipping happens for more general theories and at many more positions in the momentum space \cite{Abbasi:2019rhy,Abbasi:2020ykq,Abbasi:2020xli,Abbasi:2023myj,Ageev:2021xjk,Ahn:2019rnq,Ahn:2020bks,Ahn:2020baf,Amrahi:2023xso,Atashi:2022emh,Baishya:2023nsb,Blake:2019otz,Blake:2021hjj,Grozdanov:2018kkt,Grozdanov:2019uhi,Grozdanov:2020koi,Grozdanov:2023txs,Grozdanov:2023tag,Jansen:2020hfd,Jeong:2021zhz,Jeong:2022luo,Jeong:2023zkf,Kim:2020url,Kim:2021hqy,Kim:2021xdz,Liu:2020yaf,Mahish:2022xjz,Natsuume:2019sfp,Natsuume:2019xcy,Natsuume:2019vcv,Sil:2020jhr,Yuan:2020fvv,Yuan:2021ets,Yuan:2023tft,Wang:2022xoc,Wu:2019esr}. Moreover, even though the number of pole-skipping points at each frequency and the corresponding momenta depend on the details of the theory, there is a universal pattern: pole skipping happens at frequencies $\omega=\i(l_b-s) 2\pi T$ for all positive $s$, where $l_b$ is the highest spin in the theory. For $l_b\ge2$, we can write the leading frequency as $\omega=\i \l_L$ because $\l_L=(l_b-1)2\pi T$ is the Lyapunov exponent for a theory with a spin $l_b$ \cite{Perlmutter:2016pkf,Haehl:2018izb,Narayan:2019ove}. To derive this general statement, \cite{Wang:2022mcq} used what we will refer to as the \emph{covariant expansion formalism}. Expanding bulk fields around the black hole horizon, not using partial derivatives, $\partial_r$, but using covariant derivatives, $\nabla_r$, one finds that certain properties of the equations of motion become manifest. The covariant expansion formalism also provides an algorithm for locating skipped poles, making it automatable on a computer program.

Interestingly, this pattern of frequencies for bosonic fields was found to hold analogously in some examples with fermionic fields as well. For the theory of a minimally coupled Dirac fermion, the leading frequency was found to be $-\i \pi T$ \cite{Ceplak:2019ymw}, and, for the theory of a minimally coupled Rarita-Schwinger field, it was found to be $\i \pi T$ \cite{Ceplak:2021efc}. We will show that this pattern holds for fermionic theories in general. More explicitly, for a theory of fermions (with no gravitational backreaction) with the highest spin being $l_f$, pole skipping happens universally at frequencies $\i(l_f-s)2\pi T$ for all $s\in\mathbb{Z}^+$. 

More generally, a theory will contain both bosonic and fermionic fields that are dynamic. For example, supergravity theories have both. We will argue that there would be two towers of pole-skipping frequencies starting at $\i(l_b-1)2\pi T$ and $\i(l_f-1)2\pi T$ respectively. This is regardless of how the bosons and fermions are coupled.

The covariant expansion method relies on a classification of linearized perturbations of bulk fields and an analysis of equations of motion on this basis. In a theory with gauge symmetry, not all equations of motion are independent. This redundancy is commonly dealt with by restricting to only gauge-invariant quantities \cite{Kovtun:2005ev}, which is a theory-dependent procedure. Since part of the motivation of \cite{Wang:2022mcq} was to provide an algorithm, it is thus helpful if we have a systematic procedure for finding the pole-skipping points given a Lagrangian while sidestepping the gauge analysis. In this paper, we present a pole-skipping condition that works without having to remove the gauge redundancy, and we will refer to it as the \emph{gauge-covariant condition}.

We will begin by reviewing the covariant expansion formalism in the bosonic case in Section~\ref{sec:bosons}, followed by a discussion of gauge redundancy which leads to the gauge-covariant pole-skipping condition in Section~\ref{sec:gauge}. We then present the analogous formalism for theories with only fermions in Section~\ref{sec:fermions}. In Section~\ref{sec:gen}, we present an argument for the general pattern of pole-skipping frequencies when both bosonic and fermionic fields are present. We then discuss some consequences and future directions in Section~\ref{sec:discferm}. Some examples with bosonic fields and fermionic fields are presented in Appendices~\ref{app:egbose} and \ref{app:egfermi} respectively, where gauge symmetry is present in two of the bosonic and one of the fermionic examples.


\section{\label{sec:bosons}Bosonic fields}

We now review the covariant expansion formalism of \cite{Wang:2022mcq} which was in the context of general holographic theories with bosonic fields. Consider a local diffeomorphism-invariant action of the form
\begin{equation}\label{eq:action}
    S = \int \d^{d+2}x\,\sqrt{-g}\,\mathcal{L}\left(g,R,\nabla,\Phi\right),
\end{equation}
where $\mathcal{L}$ is constructed from contractions of the metric, $g$, the Riemann tensor, $R_{\m\n\r\s}$, bosonic matter fields which are collectively denoted as $\Phi$, and their covariant derivatives such as $\nabla_\l R_{\m\n\r\s}$ and $\nabla_\m\nabla_\n \Phi$. An example of a term that can appear in $\mathcal{L}$ is $\phi R_{\m\n\r\s} F^{\m\n}F^{\r\s}$ for some scalar field $\phi$ and some field strength $F_{\m\n}$. 

At large $N$ and finite temperature $T$, a boundary state in thermal equilibrium is described by a stationary black hole in the bulk \cite{Maldacena:2001kr}. Let us write its metric in ingoing Eddington-Finkelstein coordinates as
\begin{equation}
\label{eq:iEF}
\d s^{2}=- f(r) \d v^{2}+2 \d v \d r+h(r) \d x^i \d x^i,
\end{equation}
where $f(r_0)=0$ at the horizon $r=r_0$ and $i\in\{1,\dots,d\}$ labels the transverse directions. We will refer to it as the background metric. We can also have stationary background matter fields, which, as a simplifying assumption, are isotropic and homogeneous in $x^i$. Furthermore, all background fields should be regular at both past and future horizons. 

For future reference, the non-vanishing Christoffel symbols for our background metric are
\begin{equation}\label{eq:chris}
\begin{aligned}
\Gi^{v}_{vv}&=\frac{1}{2}f^\pr,&~~~\Gi^{v}_{ij}&=-\frac{1}{2}h^\pr\di_{ij},&~~~
\Gi^{r}_{vr}&=-\frac{1}{2}f^\pr,\\
\Gi^{i}_{rj}&=\frac{h^\pr}{2h}\di^i_j,&~~~
\Gi^{r}_{vv}&=\frac{1}{2}ff^\pr,&~~~\Gi^{r}_{ij}&=-\frac{1}{2}fh^\pr\di_{ij},
\end{aligned}
\end{equation}
and the non-vanishing components of the Riemann tensor are
\begin{equation}
\bad
R_{vrvr}&=\frac{1}{2}f^\prpr
,~~~R_{virj}=-\frac{1}{4} f^\pr h^\pr\di_{ij}
,~~~R_{rirj}=\frac{h'^2-2h h^\prpr}{4h} \di_{ij},\\
R_{vivj}&=\frac{1}{4}f f^\pr h^\pr\di_{ij},~~~R_{ijkl}=-\frac{1}{4}fh'^2\(\di_{i k}\di_{jl}-\di_{il}\di_{jk}\).
\ead
\end{equation}

Quasinormal modes are perturbations of the dynamical fields on the black hole background that satisfy the linearized equations of motion with ingoing boundary conditions at the future horizon and trivial Dirichlet boundary conditions at the asymptotic boundary, i.e., with the non-normalizable falloff set to zero \cite{Berti:2009kk,Horowitz:1999jd}. In the Fourier space with frequency denoted by $\oi$ and momenta denoted by $k_i$, finding all quasinormal modes amounts to finding a dispersion relation $\omega(k_i)$.\footnote{Because the transverse dimensions $x^i$ are non-compact, $k_i$ are continuous parameters.} The retarded Green's function, on the other hand, is computed by solving for the linearized on-shell perturbation with ingoing boundary conditions at the future horizon but non-trivial Dirichlet boundary conditions at the asymptotic boundary, i.e., with the non-normalizable mode turned on. The Green's function is then defined to be the ratio of the coefficient of the normalizable falloff to that of the non-normalizable falloff \cite{Birmingham:2001pj,Son:2002sd,Iqbal:2009fd}. It follows from these definitions that the poles of the retarded Green's function in the Fourier space are identified with the quasinormal mode spectrum, $\oi(k_i)$ \cite{Son:2002sd}. 

Recently, it has been discovered that equations of motion (along with boundary conditions at the horizon and the asymptotic infinity) do not always uniquely determine the bulk solution, leading to an ambiguity in the Green's function whose value depends on how the limit is taken in the frequency-momentum space $(\oi,k_i)$ \cite{Blake:2018leo,Blake:2019otz}. 

More precisely, at special frequencies and momenta, an equation of motion at the horizon becomes trivial. The triviality of an equation of motion implies that there is one fewer constraint than the dynamical degrees of freedom, leading to the existence of an extra ingoing mode (see \cite{Blake:2019otz} for more detailed explanations and examples). 
This unconstrained ingoing mode would then result in an ambiguity in the holographic retarded Green's function as the latter is computed from the bulk solution, now non-unique. This ambiguity was known to be related to the phenomenon of pole skipping, where the retarded Green's function at special values $\oi=\oi_*$ and $k=k_*$ ($i$-index suppressed) takes the form $G \sim 0/0$.\footnote{Technically, the ambiguity of the bulk solution is only a necessary condition for pole skipping; however, it is also sufficient in all known examples.}
To see why the Green's function is ambiguous or in other words multi-valued at pole-skipping points, expand the numerator and the denominator in terms of small $\di \oi$ and $\di k$ around such a point:
\begin{equation}
    G(\oi_*+\di \oi,k_*+\di k )\sim \frac{0+a(\di\oi)+b(\di k)}{0+c(\di\oi)+d(\di k)}\sim\frac{a(\di \oi/\di k)+b}{c(\di \oi/\di k)+d}.
\end{equation}
This illustrates the dependence of the value of $G(\oi_*,k_*)$ on the direction in which it is approached in the $(\oi,k)$ plane. In the special case $a=c=0$, called anomalous in \cite{Blake:2019otz}, the Green's function does not depend on $\di \oi/\di k$ to leading order, but the limit would now depend on higher order quantities such as $(\di k)^2$ \cite{Ahn:2020baf}. In either case, we have $0/0$, so the pole is skipped.

Pole skipping can therefore be studied by examining the equations of motion. It turns out that it is sufficient to expand the equations of motion perturbatively in the radial direction away from the future horizon $r=r_0$. This is a manifestation of the expectation that the near-horizon geometry is responsible for many universal aspects of the dual thermal system, one prime example being the near-horizon explanation \cite{Iqbal:2008by} for the universal ratio of shear viscosity to entropy density \cite{Kovtun:2004de}.

A single Fourier mode of linear perturbation of a dynamical field $X$ takes the form
\begin{equation}
\label{eq:XFourier}
\delta {X}(r, v, x)=\delta {X}(r)\, e^{-\i \omega v+\i k x},
\end{equation}
where $kx$ is a shorthand for $k_ix^i$. The Fourier coefficient $\delta {X}(r)$ still depends on the radial coordinate $r$ because the radial direction is not a coordinate of the boundary theory; since we are computing the boundary Green's function, albeit in the bulk, the radial direction should not be Fourier transformed. Often, the next step is to expand this function as a Taylor series around the horizon $r=r_0$:
\begin{equation}
    \delta {X}(r) = \sum_{n=0}^{\infty} \frac{\left.\left((\partial_r)^n\di X\right)\right|_{r=r_0}}{n!} (r-r_0)^n.
\end{equation}
However, it was noticed in \cite{Wang:2022mcq} that 
\begin{equation}\label{eq:defdiX}
    \left.\left((\nabla_r)^n \delta X\right)
    \right|_{r=r_0}, \quad n\ge 0,
\end{equation}
form a more convenient basis for the near-horizon degrees of freedom. As we will see, working with covariant expressions like these turns out to be a key idea that helps reveal various hidden symmetries of the equations of motion. We shall denote this set by $\di \mathcal{X}$; for example, $\di \mX = [\di g_{vv},\di g_{vi}, \nabla_r\di g_{vv},\dots]|_{r=r_0}$. Similarly, let us use $\di \mE=0$ to denote perturbed equations of motions and their covariant radial derivatives evaluated on the horizon. 

In order to organize the equations of motion, it is useful to define the \emph{weight} \cite{Wall:2015raa,Wang:2022mcq} for a general tensor component as
\begin{equation}\label{eq:weight_bos}
\begin{aligned}
    \text{weight}=\# (\text{lower $v$}) - \# (\text{lower $r$}) - \# (\text{upper $v$}) + \# (\text{upper $r$}),
\end{aligned}
\end{equation}
i.e., each $v$-index carries a weight of $+1$ if downstairs and $-1$ if upstairs, and the opposite is true for $r$-indices; transverse indices $i,j,\dots$ do not carry weights. With this, define $\di\mX_p$ as the subset of $\di\mX$ that has weight $p$ and similarly for $\di\mE_p$. We will think of and refer to them as vectors to compactify some of our equations, but it is nothing more than a notational convenience. For example, in Einstein gravity,
\begin{equation}
\begin{aligned}
    \di\mathcal{X}_2&=[\di g_{vv}]|_{r=r_0},\\
    \di\mathcal{X}_1&=[\nabla_r \di g_{vv},\di g_{vi}]|_{r=r_0},\\
    \di\mathcal{X}_0&=[\nabla_r\nabla_r \di g_{vv},\nabla_r \di g_{vi},\di g_{ij},\di g_{vr}]|_{r=r_0}. 
\end{aligned}
\end{equation}
More generally, with matter fields, say $B_{\m\n\r}$ and $A_{\m}$, it could contain additional terms like 
\begin{equation}
    \di\mathcal{X}_0=[\cdots,\nabla_r^3 \di B_{vvv}, \nabla_r^2 \di B_{vvi},\di B_{ijk},\di A_i, \nabla_r \di A_v,\cdots]|_{r=r_0}. 
\end{equation}
The near-horizon expansions of the perturbed equations of motion can now be compactly written as
\begin{equation}\label{eq:E=MX}
    \di \mE_p= \sum_q {\mM}_{p,q}(\omega,k)\,\di \mathcal{X}_q,
\end{equation}
where a Fourier mode of the form \eqref{eq:XFourier} has been substituted. It is worth mentioning that ${\mM}_{p,q}(\omega,k)$ for each $p$ and $q$ is a $|\di\mE_p|$-by-$|\di\mathcal{X}_q|$ matrix, the modulus sign denoting the length of the vectors. The definition of the weight has allowed us to divide the infinite matrix into these finite ones, each labeled by $p$ and $q$. This division leads to an important property that, for $p>q$,
\beq\label{eq:Mpq_factors}
\mM_{p,q}(\omega,k)\propto\[\oi-(p-1)\oi_0\]...\[\oi-q\oi_0\],
\eeq
where 
\begin{equation}
    \omega_0\equiv\i 2\pi T=\i f'(r_0)/2.
\end{equation}
The proof of this statement uses the covariance of the equation of motion and the symmetry of the background fields including in particular the metric \eqref{eq:iEF}. To see the origin of these factors, notice that a component of a linearized equation of motion is a sum of terms with the following form:
\beq
\label{eq:Fpqkl}
\di\mE_p=\sum F(g,R,\nabla,\Phi)(\nabla_v)^k(\nabla_i)^l(\nabla_r)^m\di X_{q'+m},
\eeq
where $F$ is a tensor component constructed from the background fields. To get into this form, one needs to use the Riemann tensor to commute the covariant derivatives, and it is important that $\nabla_v$ derivatives are pushed through to the very left for our purpose. 

A consequence of the background fields being stationary is that any tensor component constructed from background fields with a positive weight must vanish, so $F$ must have a non-positive weight. When $p>q'$, we must then have $k\ge p-q'$ to balance the weights on both sides. The action of $(\nabla_v)^k$ along with \eqref{eq:XFourier} then immediately leads to the factors 
\begin{equation}\label{eq:factorsqprime}
    \left[\omega-(p-1) \omega_0\right] \ldots\left[\omega-q' \omega_0\right]
\end{equation}
because they have a very simple action on a general tensor component $T$ with weight $\mathrm{w}$ when evaluated on the horizon:
\beq
\nabla_v T=\(\pd_v-\frac{\mathrm{w}}{2}f^\pr(r_0)\) T.
\eeq
It is a straightforward exercise to show this using the definition of the covariant derivative and the associated Christoffel symbols explicitly given in \eqref{eq:chris}.

To go from \eqref{eq:Fpqkl} to \eqref{eq:E=MX}, besides substituting the Fourier mode and unpacking $(\nabla_v)^k$, we still need to unpack $(\nabla_i)^l(\nabla_r)^m \delta X_{q^{\prime}+m}$, which can lead to additional terms, each proportional to $\delta \mX_q$ for some $q$. From \eqref{eq:chris}, one should notice that the only non-vanishing Christoffel symbols that can appear in $\nabla_i$ are $\Gamma_{ij}^v$ and $\Gamma_{ri}^j$, i.e.,
\begin{equation}
\label{covDi}
\begin{aligned}
    \nabla_i T_{\dots\mu\dots} &= \partial_i T + \cdots - \Gamma^\rho_{\mu i} T_{\dots\rho\dots} + \cdots\\
    &= \partial_i T + \cdots - \delta_{\mu}^j\Gamma^v_{ji} T_{\dots v\dots} - \delta_{\mu}^r\Gamma^j_{ri} T_{\dots j\dots} + \cdots,
\end{aligned}
\end{equation}
so it can at most turn a lower $i$-index into a lower $v$-index or a lower $r$ into a lower $j$, thereby increasing the weight by one in either case. The same is true for upper indices. In other words, we must have $q\ge q'$. This concludes the derivation of $\eqref{eq:Mpq_factors}$ because \eqref{eq:factorsqprime} contains at least as many factors as needed, with the extra factors playing no obvious role.

We now turn to the condition under which an equation of motion trivializes. From property \eqref{eq:Mpq_factors}, it follows that, for any integer $s>0$, matrices $\mM_{p,q}(\oi,k)$ with $p>q_0-s\ge q$ vanish identically when
\beq
\label{psomega}
\oi= (q_0-s) \, \oi_0,
\eeq
where $q_0$ is the highest possible weight of any dynamical field in the theory. Then the infinite dimension matrix
\beq\label{eq:fullmat}
\mM_\infty(\omega,k)\equiv
\begin{bmatrix}
\mM_{q_0,q_0}&\mM_{q_0,q_0-1}&...\\
\mM_{q_0-1,q_0}&\mM_{q_0-1,q_0-1}&...\\
...&...&...\\
\end{bmatrix}
\eeq
takes the form
\begin{align}
    \mM_{\infty}((q_0-s) \oi_0,k)=
    \begin{bmatrix}
       \begin{matrix}
            \phantom{0}& \phantom{0} &\phantom{0} \\
            \phantom{0} & M_s(k) &\phantom{0}\\
            \phantom{0} & \phantom{0} &\phantom{0}
        \end{matrix}
       &\vline
       \begin{matrix}
            \phantom{0}& \phantom{M_s(k)} &\phantom{0} \\
            \phantom{0} & {0} &\phantom{0}\\
            \phantom{0} & \phantom{0} &\phantom{0}
        \end{matrix}
       \\\hline
       \begin{matrix}
            \phantom{0}& \phantom{M_s(k)} &\phantom{0} \\
            \phantom{0} & {\cdots} &\phantom{0}\\
            \phantom{0} & \phantom{0} &\phantom{0}
        \end{matrix}
        & \vline
        \begin{matrix}
            \phantom{0}& \phantom{M_s(k)} &\phantom{0} \\
            \phantom{0} & {\cdots} &\phantom{0}\\
            \phantom{0} & \phantom{0} &\phantom{0}
        \end{matrix}
   \end{bmatrix}.
\end{align}
As soon as we make the finite submatrix
\beq
M_s(k)\equiv
\left.
\begin{bmatrix}
\mM_{q_0,q_0}&...&\mM_{q_0,q_0-s+1}\\
...&...&...\\
\mM_{q_0-s+1,q_0}&...&\mM_{q_0-s+1,q_0-s+1}\\
\end{bmatrix}
\right|_{\oi=(q_0-s)\oi_0}
\eeq
degenerate, a certain linear combination of the equations of motion $\di\mE_p$ with $p>q_0-s$ would become trivial, leaving a certain linear combination of the degrees of freedom $\di\mX_q$ with $q>q_0-s$ free. This generally happens at discrete values of $|k|\equiv\sqrt{k_ik_i}$. Incidentally, the number of such $|k|$ generically increases with $s$: as the size of $M_s(k)$ grows, the determinant becomes a polynomial of higher degree which generically has more roots.

To summarize, the general pole-skipping condition is given by
\beq
\label{psk}
\boxed{
    \oi=(q_0-s) \, \oi_0\quad\text{and}\quad {\rm det}~M_s(k)=0
}
\eeq
for any $s\in \mathbb{Z}^+$, where for convenience we repeat that $q_0$ is the highest weight in the theory and $\oi_0\equiv \i 2\pi T = \i f'(r_0)/2$.

\section{Gauge fields}\label{sec:gauge}

We have just reviewed the covariant expansion formalism for general holographic theories. It works well for bosonic fields without gauge symmetry. For gauge theories, the formalism works only after removing the gauge redundancy. To understand why there might be subtleties if we do not do so, realize that a pure gauge perturbation should automatically satisfy all equations of motion by their very nature. 

To see how this is reflected in the covariant expansion formalism, write
a general gauge parameter as
\begin{equation}
    \di \xi(v,r,x)=\di \xi(r) e^{-\i \omega v+\i kx},
\end{equation}
where $\xi$ could carry Lorentz indices. For example, $\di\xi(v,r,x)=\Lambda(v,r,x)$ in Maxwell theory where the gauge transformation is given by $A_\mu\to A_\mu + \nabla_\mu \Lambda$. As another example, in General Relativity, $\di\xi=\zeta_\mu$, where the gauge transformation is given by $g_{\m\n}\to g_{\m\n}+\nabla_{\m}\zeta_\n+\nabla_{\n}\zeta_\m$. With this, just like how we defined $\di \mX_{q}$ below \eqref{eq:weight_bos}, we define $\di \Xi_{u}$ to be the subset of $\nabla_r^n\di\xi|_{r=r_0}$ with weight $u$. 

With this, we can write a general pure-gauge perturbation as
\begin{align}\label{eq:ggexpn}
\di \mathcal{X}_q=\sum_u \mathcal{T}_{q,u}(\oi,k) \,\di \Xi_u.
\end{align}
Continuing with the Maxwell example, $\di A_\mu=\nabla_\m\di\xi=\nabla_\mu \L$. Writing out the components of \eqref{eq:ggexpn}, the first few orders are given by\footnote{Here, everything is evaluated on the horizon, but from now on we will frequently avoid writing $(\cdot)|_{r=r_0}$ when it is clear that the quantity should be evaluated on the horizon.}
\begin{align}
\begin{bmatrix}\di A_v \\ \di A_i\\ \nabla_r \di A_v\\ \di A_r \\ \nabla_r \di A_i \\ \nabla_r^2 \di A_v\\...\end{bmatrix}=\begin{bmatrix}-\i \oi\\ \i k_i\\ 0\\ 0 \\ -{\i k_i h^\pr}/{2h} \\ 0\\...\end{bmatrix}\Lambda+
\begin{bmatrix}0 \\ 0\\ -\i\oi+f^\pr/2\\ 1 \\ \i k_i \\ {f^\prpr}/{2}\\...\end{bmatrix} \nabla_r \Lambda+\cdots.
\end{align}
The gravitational case is very similar and will be presented in Appendix~\ref{Einsteingravity}.

Once again, $\mathcal{T}_{q,u}$ satisfies the property of being proportional to $\left[\omega-(q-1) \omega_0\right]$ $\ldots$ $\left[\omega-u \omega_0\right]$ for $q>u$ for exactly the same reason, namely that we need sufficiently many $\nabla_v$ to raise the weight from $u$ to $q$, as reviewed in Section~\ref{sec:bosons}.

A pure-gauge perturbation has the property that it automatically satisfies the equations of motion
\begin{align}\label{eq:gaugeinf}
\di \mathcal{E}_p=\sum_{q,u}\mathcal{M}_{p,q}\,\mathcal{T}_{q,u}\,\di \Xi_u=0\quad \forall p.
\end{align}
This can help us understand some features of the matrices $\mathcal{M}_{p,q}$. 
At $\oi=(q_0-s)\oi_0$ where $s>0$, as explained earlier, the entries of $\mathcal{M}_{p,q}$ with $p>q_0-s\geq q$ will be zero. At the same time, for the same reason, the entries of $\mathcal{T}_{q,u}$ with $q>q_0-s\ge u$ will be zero. Therefore, the infinite-dimensional statement \eqref{eq:gaugeinf} reduces to a finite one:
\begin{align}\label{eq:gaugefinite}
&\sum_{q=q_0-s+1}^{q_0} \sum_{u=q_0-s+1}^{u_0}\mathcal{M}_{p,q}\,\mathcal{T}_{q,u} \,\di \Xi_u=0\nn
\Rightarrow \, &\sum_{q=q_0-s+1}^{q_0}\mathcal{M}_{p,q}\(\mathcal{T}_{q,u_0} \di \Xi_{u_0}+...+\mathcal{T}_{q,q_0-s+1} \di \Xi_{q_0-s+1}\)=0,~~~q_0-s+1\leq p\leq q_0.
\end{align}
Since $\di\Xi_u$ is the gauge parameter, we can choose its value for different $u$ independently. Furthermore, for a given $u$, we can even choose its $|\di\Xi_u|$ entries independently. From \eqref{eq:gaugefinite}, it is now clear that, for $\di\Xi_u$ with a given $u>q_0-s$, each $\mathcal{T}_{q,u}\di\Xi_u$ (with the range of $q$ restricted to $q_0-s+1\leq q\leq q_0$) belongs to the kernel of $M_s(k)$.

Before moving on, let us comment on the visibility of gauge redundancy at different orders. Notice that $u_0< q_0$ usually. For example, in Maxwell theory, the highest weight for $\di\mX$ is $q_0=1$ owing to $A_v$, whereas the highest weight for $\di\xi=\Lambda$ is $u_0=0$ as it is a scalar; similarly, in Einstein or higher-derivative gravity, $q_0=2$ owing to $\di g_{vv}$, but $u_0=1$ because $\di \xi=\zeta_\mu$ is a vector. This means that, in both cases, the matrix $M_s(k)$ has no kernel at leading order ($s=1$). This is due to the fact that the range of $u$-index in \eqref{eq:gaugefinite} is empty if $u_0<q_0-s+1$. As soon as $s$ is large enough, however, a kernel will exist for all larger values of $s$. 

In summary, the determinant of $M_s(k)$ will always vanish automatically, i.e., without having to fine-tune $k$, for all $s\ge q_0-u_0+1$. This invalidates our earlier proposal for the pole-skipping condition \eqref{psk} as we expect to turn $\det M_s(k)$ to zero only at special $k$. 

Here is the moral of the story. This automatic degeneracy, as we have just seen, is a manifestation of gauge redundancy, i.e., gauge symmetry makes some equations of motion redundant. Pole skipping, however, is a statement about the physical bulk solution having an extra degree of freedom. Therefore, if we want to stick to \eqref{psk} as our pole-skipping condition, we would have to remove the redundancy at the onset. This can be done by e.g.~restricting to the physical (gauge-invariant) degrees of freedom and their equations of motion as in \cite{Kovtun:2005ev}.

In practice, we find it convenient to skip the step of figuring out the gauge symmetry of the theory and finding the physical degrees of freedom. This is particularly advantageous if whether a theory has gauge symmetry depends on the values of certain coupling constants of the theory. For example, for a theory of Rarita-Schwinger field on curved space with mass $m$, there is gauge symmetry if $m=d/2$ and the curved background satisfies vacuum Einstein's equation with a negative cosmological constant, and no gauge symmetry otherwise. We study this example in Appendix~\ref{RSfield}. 

With the understanding above, we now present what we call the gauge-covariant version of \eqref{psk}:
\begin{equation}\label{psknew}
\boxed{
    \omega = (q_0-s)\, \oi_0\quad \text{and}\quad \dim(\ker M_s(k))\nearrow
}
\end{equation}
for each $s\ge1$. To state it in words, for a given pole-skipping frequency, pole skipping happens for values of $k$ that increase the dimension of the kernel (from its dimension at generic values of $k$).\footnote{Incidentally, increasing the dimension of the kernel is different from setting non-zero diagonal entries of the Jordan normal form to zero. As an example, the matrix
\begin{align}
\begin{bmatrix}
0&\phantom{0}1
\\0&\phantom{0}x
\end{bmatrix}
\end{align}
has eigenvalues $0$ and $x$, but setting $x=0$ will not expand its kernel from one dimension to two. Terminology-wise, increasing the algebraic multiplicity does not necessarily increase the geometric multiplicity. The geometric multiplicity of an eigenvalue is the dimension of its eigenspace. The algebraic multiplicity of an eigenvalue is its multiplicity as a root of the characteristic polynomial $\det (\li \mathbbm{1} - M)$ for the matrix $M$. A given eigenvalue's algebraic multiplicity is equal to or greater than its geometric multiplicity. The kernel's dimension is the geometric multiplicity of the eigenvalue zero.\label{ft:jordan}
}

As examples, we use this new prescription to study Maxwell theory in Appendix~\ref{Maxwelltheory}, Einstein gravity in Appendix~\ref{Einsteingravity}, and Rarita-Schwinger theory in Appendix~\ref{RSfield}. For other examples where there is no gauge symmetry, the condition \eqref{psknew} reduces to \eqref{psk}.

\section{Fermionic fields}\label{sec:fermions}

Let us now turn to bulk theories with only fermionic degrees of freedom. In this case, bosonic fields (including the metric) exist only as fixed backgrounds. This assumption is justified for example in theories with minimally coupled fermionic matter where the matter action carries an extra factor of $G_N$ relative to the gravitational action since the gravitational backreaction can be neglected at leading order in $G_N$. Alternatively, if the fermionic background is trivial, the stress tensor remains zero at linear order, so we can neglect the backreaction in this case, too. With only dynamical fermionic bulk fields, the resulting pole skipping will be for holographic Green's functions of boundary fermionic operators, so the results in this section also help understand the analytic structure of such boundary quantities. Another motivation for this section is to demonstrate how the techniques used in the bosonic case generalize, as many tools from this section will be useful when we consider general theories with both types of dynamical fields in the next section. 

To begin with, the metric is again given by \eqref{eq:iEF}, repeated here for readability: 
\begin{align}
\d s^2=-f(r) \d v^2+2\d v\d r+h(r) \d x^i \d x^i.
\end{align}
The transverse direction $x^1$ is now $x$ for notational simplicity. 
To couple to fermions on a curved background, we need to introduce the tetrad, or frame fields. Just like how a useful choice of coordinates made various properties of the metric manifest, a specific choice for the tetrad will be similarly useful. Taking $(\bv,\br,\bx^i)$ as coordinates for the ($d+2$)-dimensional Minkowski spacetime with
\begin{equation}
    \ei_{\bv\bv}=-1, \quad\ei_{\br\br}=1, \quad \ei_{\bar{\imath}\bar{\jmath}}=\delta_{\bar{\imath}\bar{\jmath}},
\end{equation}
we choose the frame fields to have components
\begin{align}
e_v^\bv=\frac{1}{2}\(1+f\),~~~e_r^\bv=-1,~~~e_v^\br=\frac{1}{2}\(1-f\),~~~e_r^\br=1,~~~e_i^{\bar{\jmath}} =\sqrt{h} \,\di_i^{\bar{\jmath}},
\end{align}
which satisfy $g_{\m\n}=\eta_{{a}{b}}e^{a}_{\m} e^{b}_{\n}$. Bars over the flat space indices are used to distinguish them from the curved spacetime indices, and we use the Latin alphabet $a,b,\dots$ to denote them abstractly. 

The associated non-vanishing spin connections are then given by
\begin{align}
\(\oi_{\bv\br}\)_v=-\frac{1}{2}f^\pr,~~~\(\oi_{\bv\bar{\imath}}\)_j=\frac{1}{4}\frac{h^\pr}{\sqrt{h}}\(1-f\)\di_{\bar{\imath}j},~~~\(\oi_{\br\bar{\imath}}\)_j=-\frac{1}{4}\frac{h^\pr}{\sqrt{h}}\(1+f\)\di_{\bar{\imath} j}.
\end{align}
Note that the barred indices are anti-symmetric. We also use $\nabla_\mu$ to denote the full covariant derivative, whose action depends on the object it acts on. For example, for a spinor $\psi$, 
\begin{equation}
    \nabla_\mu\psi=\partial_\mu\psi+\frac{1}{4}(\oi_{{a}{b}})_\mu\Gamma^{{a}{b}}\psi,
\end{equation}
and for a vector-spinor $\psi_\mu$,
\begin{equation}
    \nabla_\mu\psi_\nu=\partial_\mu\psi_\nu+\frac{1}{4}(\oi_{{a}{b}})_\mu\Gamma^{{a}{b}}\psi_\nu-\Gamma^\rho_{\mu\nu}\psi_\rho.
\end{equation}

To avoid unnecessary complications caused by the dimension-dependent nature of gamma matrices, we will from now on work in $2+1$ bulk dimensions; the generalization to higher dimensions is straightforward and will be discussed briefly at the end of the section. With $d=1$, we have the following three gamma matrices:
\begin{align}\label{3Dgammamatrices}
\Gi^\bv=
    \begin{bmatrix}
        \phantom{0}0&\phantom{0}1\\
        -1&\phantom{0}0
    \end{bmatrix},~~~
\Gi^\br=
    \begin{bmatrix}
    0&\phantom{0}1\\1&\phantom{0}0
    \end{bmatrix},~~~
    \Gi^\bx=
    \begin{bmatrix}-1&\phantom{0}0\\
    \phantom{0}0&\phantom{0}1\end{bmatrix}.
\end{align}
In curved spacetime coordinates,
\begin{align}
\Gi^v=\begin{bmatrix}0&\phantom{0}2\\
0&\phantom{0}0\end{bmatrix},
\quad
\Gi^r=\begin{bmatrix}0&\phantom{0}f\\1&\phantom{0}0\end{bmatrix},
\quad
&\Gi^x=\frac{1}{\sqrt{h}}\begin{bmatrix}-1&\phantom{0}0\\\phantom{0}0&\phantom{0}1\end{bmatrix}.
\end{align}
Define projectors
\begin{align}
P_\pm=\frac{1 \mp \sqrt{h}\Gi^x}{2}=\frac{1 \mp\Gi^{\bar x}}{2},
\end{align}
which, for a general fermionic quantity $X$ (potentially carrying Lorentz indices), has the following effect:
\begin{align}
X=\begin{bmatrix}X_+\\X_-\end{bmatrix},
\quad
P_+X=\begin{bmatrix}X_+\\0\end{bmatrix},
\quad
P_-X=\begin{bmatrix}0\\X_-
\end{bmatrix}.
\end{align} 
Just like how we decomposed Lorentzian tensors into its components when considering bosons, we do the analogous thing here of decomposing all fermionic quantities into $\pm$ components. 
For example, they could be $\di \psi_+, \nabla_r\di \psi_-, \nabla_r\nabla_r \di \Psi_{v,+}$, etc. 
An operator acting on a spinor carries two spinor indices. For such objects, we decompose in the same way and write
\begin{equation}
    O=\begin{bmatrix}O_+{}^+&O_+{}^-\\
    O_{-}{}^+&O_{-}{}^-\end{bmatrix}.
\end{equation}

This allows us to easily generalize the definition of the weight. We define a lower $\pm$ to carry $\pm1/2$ weight and an upper $\pm$ to carry $\mp1/2$ in addition to the contributions from Lorentz indices, i.e., the total weight is given by
\begin{align}\label{eq:weightferm}
    \text{weight}=\,&\# (\text{lower $v$})&&- \# (\text{lower $r$})& &- \# (\text{upper $v$}) &&+ \# (\text{upper $r$})
    \nn
    +\,& \frac{1}{2}\# (\text{lower +})&&-\frac{1}{2} \# (\text{lower $-$}) &&-\frac{1}{2} \# (\text{upper $+$}) &&+\frac{1}{2} \# (\text{upper $-$}).
\end{align}
The analogue of \eqref{eq:E=MX} can be written as
\begin{equation}\label{eq:E=MXferm}
\delta \mathcal{E}_p=\sum_q \mathcal{M}_{p, q}(\omega, k)\, \delta \mathcal{X}_q,
\end{equation}
which looks exactly the same, but $p$ and $q$ are now half-integers. Each entry in $\delta \mathcal{E}_p$ or $\delta \mathcal{X}_q$ is not just a tensor component, but a tensor-spinor projected onto one of the two eigenspaces of $\Gamma^x$. As before, we absorb $\nabla_r$ derivatives into the definition of $\di \mX$, i.e., $\nabla_r$ should not be unpacked. In the bosonic case, we have explained that $(\nabla_r)^n\delta X|_{r=r_0}$ form a better basis than partial derivatives. Here, we are going one step further by saying that
\begin{equation}
        \left.(\nabla_r)^n \begin{bmatrix}\di X_+
        \\
        \di X_-
        \end{bmatrix}
        \right|_{r=r_0}
        =
        \begin{bmatrix}((\nabla_r)^n \di X)_+|_{r=r_0}\\((\nabla_r)^n \delta X)_-|_{r=r_0}\end{bmatrix}
\end{equation}
is a good basis for packaging things.

Recall that the most important property of the matrix $\mathcal{M}_{p, q}(\omega, k)$ is \eqref{eq:Mpq_factors}. We now proceed to show that it holds for fermions as well. A general component of the equation of motion for a fermion can be written as
\begin{align}\label{eq:fermionEOMgen}
    (\nabla_r)^n\di \begin{bmatrix}
    {E}_+\\ {E}_-
    \end{bmatrix}
    =\sum
    F(g,R,\nabla,\Phi,\Psi)
    \(\nabla_v\)^k\(\nabla_i\)^l (\nabla_r)^m \di \begin{bmatrix}
    X_+\\X_-
    \end{bmatrix},
\end{align}
where $F$ is a component of a spacetime tensor and at the same time a spinor operator, and the sum is over different terms of this form. Because the three gamma matrices together with the identity matrix form a complete basis for all 2-by-2 matrices and because $F$ is itself a component of a covariant tensor, we can decompose it into
\begin{align}\label{eq:Fexpand}
    F(g,R,\nabla,\Phi,\Psi)  =V_\mu(g,R,\nabla,\Phi,\Psi) \Gamma^\mu+F_0 (g,R,\nabla,\Phi,\Psi)\mathbbm{1},
\end{align}
where $V_{\mu}$ is a vector and $F_0$ is a scalar, neither of which carries spinor indices. 


The fermionic analogue of \eqref{eq:Fpqkl} can be written as
\begin{equation}\label{eq:EFDDDX}
\delta \mathcal{E}_p =\sum \mathcal{F}(g, R, \nabla, \Phi,\Psi)\left(\nabla_v\right)^k\left(\nabla_i\right)^l\left(\nabla_r\right)^m \delta X_{q^{\prime}+m}.
\end{equation}
What is different from \eqref{eq:Fpqkl} is that $\delta \mE$ and $\di X$ both carry a $\pm$ index, and $\mathcal{F}$, being a projected component of an operator in the spinor space, carries two of them. We emphasize again that $\nabla_\mu$ here contains spin connections, so they are also operators on the spinors.

Like in the bosonic case, we first need to show that $\mathcal{F}$ vanishes on the horizon if it has a positive weight. In addition to bosonic constituents, fermionic fields are also potential ingredients. The trick here is to realize that $\mathcal{F}$ is a component of the matrix $V_\mu\Gamma^\mu+F_0\mathbbm{1}$. Since $V_\mu$ and $F_0$ do not carry spinor indices, the boost symmetry argument in the bosonic case applies still. Following our definition of the weight in \eqref{eq:weightferm}, it is straightforward to check that $\Gi^\mu$ and $\mathbbm{1}$ all have this property. For example, the only entry in the identity matrix that carries a positive weight is the ${(\mathbbm{1})}_+{}^-$ component, which is zero indeed; the components of 
\begin{equation}
    \Gi^r=\begin{bmatrix}(\Gi^r)_+{}^+&(\Gi^r)_+{}^-\\(\Gi^r)_-{}^+&(\Gi^r)_-{}^-\end{bmatrix}=\begin{bmatrix}0&\phantom{0}f\\1&\phantom{0}0\end{bmatrix}
\end{equation}
have weights $1,2,0$ and $1$, so the only one that does not have to vanish on the horizon is $(\Gamma^r)_-{}^+$; for
\be
\Gi^v=\begin{bmatrix}0&\phantom{0}2\\0&\phantom{0}0\end{bmatrix},
\ee
the components all have non-positive weights, so there is no requirement for any component to be zero even though some of them are. Since $\mathcal{F}$ is built from $V_\mu$, $F_0$, $\Gi$ and $\mathbbm{1}$, the fact that $\mathcal{F}=0$ in \eqref{eq:EFDDDX} if it has a positive weight is now guaranteed. 

To show \eqref{eq:Mpq_factors}, we just need to look at $\nabla_v$, which is the only way to increase the weight, knowing that $\mathcal{F}$ cannot. Consider a spinor $\psi$ first:
\begin{align}
    \nabla_v \begin{bmatrix}\psi_+\\\psi_-\end{bmatrix}=\begin{bmatrix}(\pd_v-f^\pr/4)\psi_+\\(\pd_v+f^\pr/4)\psi_-\end{bmatrix}.
\end{align}
This clearly satisfies 
\begin{equation}
\nabla_v T \propto\left(\partial_v-\frac{\mathrm{w}}{2} f^{\prime}\left(r_0\right)\right) T
\end{equation}
since $\psi_\pm$ have weights $\pm1/2$. For a more general tensor-spinor, we just need to take into account Christoffel symbols in $\nabla_v$, but those were exactly the same as for bosons! This concludes the proof that $\mathcal{F}$ is proportional to \eqref{eq:factorsqprime}. 

To show that $\mathcal{M}_{p,q}$ is proportional to \eqref{eq:Mpq_factors}, which is what we really need, we must show $q\ge q'$, where $q'$ is the weight of $(\nabla_r)^m\delta X_{q'+m}$ appearing in \eqref{eq:EFDDDX} and $q$ is what appears in \eqref{eq:E=MXferm}. They differ because $\nabla$'s will need to operate on the quantities to their right before evaluating everything on the horizon. In addition to Christoffel symbols, spin connections also appear in this process. Notice that $\nabla_i$ acts on a spinor as
\begin{align}
&\(\nabla_i \di\psi\)_+
=\pd_x\di \psi_+-\frac{h^\pr f}{4\sqrt{h}}\di \psi_-,
\\
&\(\nabla_i \di\psi\)_-=\pd_x \di \psi_- + \frac{h^\pr}{4\sqrt{h}}\di \psi_+.
\end{align}
Combined with \eqref{covDi}, we see that $\nabla_i$ will only turn a tensor-spinor component into another tensor-spinor component with a higher weight (when evaluated at the horizon). This ensures that $q\ge q'$. Finally, substitution of the Fourier expansion gives the desired factors \eqref{eq:Mpq_factors}. 

In fact, this concludes the discussion of bulk theories with only dynamical fermions (in three dimensions), because the rest follows in exactly the same way as in the bosonic case. The only difference is that $p,q$ are half-integers. Even the conclusion reads the same as before: pole skipping happens at frequencies $(q_0-s)\,\oi_0$ for $s=1,2,\dots$, where $q_0$ is the highest weight present in the theory, now a half-integer. 

So far, we have not said anything about gauge symmetry. Fortunately, this goes through just like in the bosonic case. Without gauge symmetry, the pole-skipping condition is given in \eqref{psk}; with gauge symmetry, we could either restrict to gauge-invariant quantities and use the same condition, or we can use the gauge-covariant condition \eqref{psknew}. We note, however, that the gauge parameter $\di\Xi_u$ appearing in \eqref{eq:ggexpn} would have to become a fermionic one, i.e., $u$ should take half-integer values. Having a bosonic gauge parameter in a theory of only dynamical fermions is neither common nor within the scope of the current section, but it does belong to the more general class we study in Section~\ref{sec:gen}.

\subsection*{Higher dimensions}

To define a spinor on curved spacetime, we need to define the gamma matrices in Minkowski space first. The gamma matrices in Minkowski space $\mathbb{R}^{1,D-1}$ satisfy the Clifford algebra $C\ell(1,D-1)$: $\{\gi^a,\gi^b\}=2 \ei^{ab}$. We can define these gamma matrices recursively, starting from two dimensions, where we can choose
\begin{align}
\gi_2^0=\begin{bmatrix}\phantom{0}0&\phantom{0}1\\-1&\phantom{0}0\end{bmatrix},~~~\gi_2^1=\begin{bmatrix}0&\phantom{0}1\\1&\phantom{0}0\end{bmatrix}.
\end{align}
The $(2n+1)$-dimensional gamma matrices are then defined using the $(2n)$-dimensional gamma matrices by
\begin{align}
&\gi_{2n+1}^a=\gi_{2n}^a,~~~\qquad\qquad a = 0,\dots,2n-1,\nn
&\gi_{2n+1}^{2n}=\i^{n+1}\gi_{2n}^0...\gi_{2n}^{2n-1}.
\end{align}
Similarly, the $(2n)$-dimensional gamma matrices are defined from $(2n-1)$-dimensional gamma matrices by
\begin{align}
&\gi_{2n}^a=\gi_{2n-1}^a\otimes \begin{bmatrix}1&\phantom{0}0\\0&-1\end{bmatrix},~~~\quad a = 0,\dots,2n-2,\nn
&\gi_{2n}^{2n-1}=\mathbbm{1}\otimes \begin{bmatrix}0&\phantom{0}1\\1&\phantom{0}0\end{bmatrix}.
\end{align}
It is then straightforward to check that these matrices satisfy the Clifford algebra. In this construction, gamma matrices are $2^{\lfloor\frac{D}{2}\rfloor}\times2^{\lfloor\frac{D}{2}\rfloor}$ matrices.


Now we can define the $\Gi$-matrices in the ($\bar{v}$, $\bar{r}$, $\bar{x}^i$) coordinates by
\begin{align}
\Gi_D^\bv&=\gi_D^0,\qquad\Gi_D^\br=\gi_D^1,\nn
\Gi_D^{\bar{\imath}}
&=\gi_D^{i}\delta_{\bar{\imath}, i-1}, \qquad \bar{\imath}=1,\dots,d.
\end{align}
For $D=3$ ($d=1$), this reproduces \eqref{3Dgammamatrices}.

The projectors to the subspaces are defined by
\begin{align}
P_+=\frac{1+\Gi^\bv \Gi^\br}{2}=\begin{bmatrix}1\\0\end{bmatrix}
\otimes \mathbbm{1}_{2^{\lfloor{D/2-1}\rfloor}},~~~P_-=\frac{1-\Gi^\bv \Gi^\br}{2}=\begin{bmatrix}0\\1\end{bmatrix}
\otimes \mathbbm{1}_{2^{\lfloor{D/2-1}\rfloor}}.
\end{align}
Here, even though each subspace would have more degrees of freedom than in $D=3$ (e.g., $\psi_\pm$ each has $2^{\lfloor{D/2-1}\rfloor}$ components), in the basis we have defined, there is no need to distinguish them further as in \cite{Ceplak:2019ymw,Ceplak:2021efc}. We attribute this difference to the choice of projectors. 

Using these $\Gi$-matrices, most of the argument for $D=3$ goes through in the same way, but some aspects must be generalized. For example, instead of having $\Gamma^\mu$ and $\mathbbm{1}$ in \eqref{eq:Fexpand} as a complete basis for the general operator $F$, a complete basis in higher dimensions is formed using $\mathbbm{1}$ and commutators of all the gamma matrices \cite{Polchinski:1998rr}. Explicitly, in even dimensions ($D=2k$), a complete basis for $2^k\times 2^k$ matrices is given by 
\begin{align}
\mathbbm{1}\cup\left\{\Gi^{[\mu_1}\Gi^{\mu_2}...\Gi^{\mu_m]} \mid m=1,..., 2k \right\},
\end{align}
and in odd dimensions ($D=2k+1$), a complete basis for $2^k\times 2^k$ matrices is given by
\begin{align}
\mathbbm{1}\cup\left\{\Gi^{[\mu_1}\Gi^{\mu_2}...\Gi^{\mu_m]} \mid m=1,..., k \right\}.
\end{align}

\section{Bosonic and fermionic fields}\label{sec:gen}
We discussed general theories with either dynamical bosonic fields or fermionic fields in earlier sections. It is then natural to ask whether the argument generalizes to the case when both are present. A naive expectation might be that, if we have already worked out the pole-skipping points for a theory with e.g. only bosonic fields, adding fermions will not change them even though it may lead to more. This is not always correct because fermions can appear even in the bosonic equations of motion, adding extra terms proportional to fermionic perturbations, so the special frequencies that could turn the original bosonic equations of motion trivial would no longer necessarily do so. In other words, the linearized bosonic equations of motion $\di\mE_B$ takes the form
\begin{equation}
    \delta \mE_B = \mathcal{M}_{BB} \delta \mX_{B} +  \mathcal{M}_{BF}  \delta \mX_{F},
\end{equation}
where $\di\mX_B$ and $\di\mX_F$ are bosonic and fermionic perturbations respectively, so even if special frequencies and momenta set $\mathcal{M}_{BB}$ to zero, $\mathcal{M}_{BF}$ can still be non-zero, preventing a linear combination of $\delta \mE_B$ from necessarily becoming trivial. 

In this section, we will consider this general case. To begin with, write covariant expansion coefficients of the equations of motion evaluated on the horizon as
\begin{equation}\label{eq:E=MXgen}
    \di \mE_p= \sum_q {\mM}_{p,q}(\omega,k)\,\di \mathcal{X}_q.
\end{equation}
This looks exactly like \eqref{eq:E=MX} and \eqref{eq:E=MXferm}, but $p,q$ can now both be integers or half-integers. When $p$ is an integer, this is a bosonic equation of motion, receiving contributions from both bosonic field perturbations which have integer $q$'s and fermionic field perturbations which have half-integer $q$'s; when $p$ is a half-integer, this is a fermionic equation of motion, again receiving contributions from both integer and half-integer $q$'s. 

Section~\ref{sec:gauge} presented a gauge-covariant formalism. It would be natural if we now proceed with the current section gauge-covariantly. However, as we will see, when both bosons and fermions are present, it is not obvious whether there exists any systematic and practical way of locating the pole-skipping momenta $k$. Nevertheless, we can still derive a general pattern of pole-skipping frequencies $\omega$. Since the main motivation for the gauge-covariant formalism was to compute the pole-skipping momenta with less effort, its advantage is lost if we are uncommitted to that goal. As a result, we find it easier to derive our statements after removing gauge redundancy. We will comment more on this issue at the end of the section.

In this approach, we still write \eqref{eq:E=MXgen}, now with the understanding that only physical (gauge-invariant) degrees of freedom and their corresponding equations of motion are included. Let us also organize the basis so that the matrix $\mM_\infty$ defined in \eqref{eq:fullmat} divides into four blocks of infinite size:
\begin{align}\label{eq:MforBFbig}
        \mM_\infty = 
    \begin{bmatrix}
  \begin{matrix}
  \mM_{l_b,l_b} & \mM_{l_b,l_b-1} &\cdots \\
  \mM_{l_b-1,l_b} & \mM_{l_b-1,l_b-1} &\cdots\\
  \cdots  & \cdots &\cdots
  \end{matrix}
  & \vline & 
    \begin{matrix}
  \mM_{l_b,l_f} & \mM_{l_b,l_f-1} &\cdots \\
  \mM_{l_b-1,l_f} & \mM_{l_b-1,l_f-1} &\cdots\\
  \cdots  & \cdots &\cdots
  \end{matrix}
  \\
    \hline
  \begin{matrix}
  \mM_{l_f,l_b} & \mM_{l_f,l_b-1} &\cdots \\
  \mM_{l_f-1,l_b} & \mM_{l_f-1,l_b-1} &\cdots\\
  \cdots  & \cdots &\cdots
  \end{matrix}
  & \vline &
  \begin{matrix}
  \mM_{l_f,l_f} & \mM_{l_f,l_f-1} &\cdots \\
  \mM_{l_f-1,l_f} & \mM_{l_f-1,l_f-1} &\cdots\\
  \cdots  & \cdots &\cdots
  \end{matrix}
    \end{bmatrix},
\end{align}
where $l_b$ is the highest integer weight and $l_f$ is the highest half-integer weight. With this separation, we can give each of the blocks a name so that $\di\mE=\mM_\infty \di\mX$ can be written as 
\begin{align}\label{eq:E=MXblocks}
    \begin{bmatrix}
        \phantom{0}\\
       \di\mE_B\\
       \phantom{0}\\
       \hline
       \phantom{0}\\
       \di\mE_F\\
       \phantom{0}
   \end{bmatrix}
   = \begin{bmatrix}
       \begin{matrix}
            \phantom{0}& \phantom{0} &\phantom{0} \\
            \phantom{0} & \mM_{BB} &\phantom{0}\\
            \phantom{0} & \phantom{0} &\phantom{0}
        \end{matrix}
       &\vline
       \begin{matrix}
            \phantom{0}& \phantom{0} &\phantom{0} \\
            \phantom{0} & \mM_{BF} &\phantom{0}\\
            \phantom{0} & \phantom{0} &\phantom{0}
        \end{matrix}
       \\\hline
       \begin{matrix}
            \phantom{0}& \phantom{0} &\phantom{0} \\
            \phantom{0} & \mM_{FB} &\phantom{0}\\
            \phantom{0} & \phantom{0} &\phantom{0}
        \end{matrix}
        &\vline 
        \begin{matrix}
            \phantom{0}& \phantom{0} &\phantom{0} \\
            \phantom{0} & \mM_{FF} &\phantom{0}\\
            \phantom{0} & \phantom{0} &\phantom{0}
        \end{matrix}
   \end{bmatrix}
    \begin{bmatrix}
        \phantom{0}\\
       \di\mX_B\\
       \phantom{0}\\
       \hline
       \phantom{0}\\
       \di\mX_F\\
       \phantom{0}
   \end{bmatrix}.
\end{align}
Recall that pole skipping happens when $\mM_{\infty}$ is degenerate. When we only have bosonic fields, we only have $\mM_{BB}$; when we only have fermionic fields, we only have $\mM_{FF}$. The problem with $\mM_{BF}$ and $\mM_{BB}$ is that they interpolate between a half-integer weight and an integer weight, so we do not have a relation like \eqref{eq:Mpq_factors}. (The argument does not generalize to this case.) 

As is by now clear, a main theme of the covariant expansion formalism is to reorganize things to manifest hidden features, so it is what we will now do once again. Acting on both sides of \eqref{eq:E=MXgen} with the invertible matrix
\begin{align}
    \mathcal{U}\equiv
    \begin{bmatrix}
       \begin{matrix}
            \phantom{0}&\mathbbm{1} &\phantom{0}\\
            \phantom{0}&\phantom{0} &\phantom{0}
        \end{matrix}
       &
       \begin{matrix}
            \phantom{0} & -\mM_{BF}\mM^{-1}_{FF} &\phantom{0}\\
            \phantom{0} & \phantom{0} &\phantom{0}
        \end{matrix}
       \\
       \begin{matrix}
            \phantom{0} & 0&\phantom{0}\\
        \end{matrix}
        &
        \begin{matrix}
            \phantom{0} & \mathbbm{1} &\phantom{0}\\
        \end{matrix}
   \end{bmatrix}
\end{align}
defines a new basis for $\di\mE$ while keeping the same basis for $\di \mX$. This physically means that we are taking linear combinations of the original equations of motion. In this basis, \eqref{eq:E=MXgen} becomes
\begin{align}\label{eq:Uaction}
    \mathcal{U}\di \mE 
    &= \mathcal{U}\mM_{\infty} \di \mX
    \nn
    \[\begin{array}{c}
         \di \mE_B - \mM_{BF}\mM_{FF}^{-1}\di\mE_F  \\[3mm]
         \di \mE_F\\
    \end{array}\]
    &= 
    \[
    \begin{array}{cc}
       \mM_{BB}-\mM_{BF}\mM_{FF}^{-1}\mM_{FB}  & 0 \\[3mm]
       \mM_{FB}  & \mM_{FF}
    \end{array}
    \]
    \[\begin{array}{c}
         \di\mX_B  \\[3mm]
         \di \mX_F
    \end{array}\].
\end{align}
With the top-right block set to zero, we have now reduced the problem of finding the conditions under which an equation of motion in the upper-left block becomes trivial. But this is the same problem as in the bosonic case! We conclude that pole skipping happens at frequencies $\i(l_b-s)2\pi T$ for $s\in\mathbb{Z}^+$. Similarly, we can use another invertible matrix to turn the lower-left block to zero, which immediately leads to the conclusion that pole skipping \emph{also} happens at frequencies $\i(l_f-s)2\pi T$ for $s\in\mathbb{Z}^+$. This concludes the proof.

We should note that this proof uses the inverse of an infinite-dimensional matrix $\mM_{FF}$ to eliminate the whole upper-right block of the equation of motion matrix. The inverse should exist after removing gauge redundancy because all equations of motion are linearly independent. It would be good to prove this rigorously. The existence of the inverse is certainly sufficient for the next steps, but it may not be necessary. In particular, when we look for pole skipping by studying \eqref{eq:Uaction}, we only need finitely many rows of the upper-right block of $\mathcal{U}\mM_\infty$ to vanish at each order in $s$. This suggests that it might be possible to remove the need to invert an infinite matrix from the argument. 

To go from the current approach to a gauge-covariant approach is in principle easy. The inverse of $\mM_{FF}$ might not exist with gauge symmetry.\footnote{It certainly would not exist if the gauge parameter is fermionic and only acts on the fermions. Similarly the inverse of $\mM_{BB}$ would not exist if the gauge parameter is bosonic and only acts on the bosons. This follows from the discussion in Section~\ref{sec:gauge}.} The physical reason why we need to invert $\mM_{FF}$ is because we want to use the fermionic equations of motion to turn fermionic perturbations into bosonic ones. This procedure can be of course performed in the gauge-covariant approach, but instead of inverting the whole matrix, we should only invert the ``physical part" of the matrix. More precisely, we should project out the kernel before performing the inverse. We have avoided doing that to keep the equations simple.

\section{Discussion}\label{sec:discferm}

In this paper, we studied pole skipping in the presence of gauge and fermionic fields. 

In the presence of gauge fields, we presented a pole-skipping condition that automatically deals with gauge symmetry. This upgrade is in fact quite practical as it allows one to compute pole-skipping points for any theory with a given Lagrangian without having to worry about removing the redundancy which usually involves determining the gauge-invariant combinations of field components. This condition reduces to the one in \cite{Wang:2022mcq} in the absence of gauge symmetry.

For theories with only fermionic fields, we provided a formalism that is parallel to the bosonic case. This formalism is again practically useful and allows one to locate the pole-skipping points systematically. With this extension, we found that pole skipping generally happens at frequencies $\i (l_f-s)2\pi T$ for positive integer $s$, where $l_f$ is the highest spin in the fermionic theory. 

We then applied the formalism to theories with both dynamic bosonic and fermionic fields. In this case, we provided an argument for the pole-skipping frequencies, namely that there is one tower of pole skipping at frequencies $\i (l_b-s)2\pi T$ and another tower with $\i (l_f-s)2\pi T$. This statement is nontrivial as the presence of fermionic fields highly influences the pole-skipping momenta of the bosonic tower and vice versa. Unlike the purely bosonic or fermionic cases, the argument here is rather abstract -- it might be difficult to actually do the computations for a specific theory due to the need to invert an infinite matrix. It would be interesting to see if this requirement can be lifted. There is one situation, however, where we do not need this inverse even when both dynamical bosons and fermions are present: when the background fermions are all zero, the bosonic and fermionic perturbations then decouple ($\mM_{BF}=\mM_{FB}=0$ in \eqref{eq:E=MXblocks}), so the problem becomes equivalent to a purely bosonic theory plus a purely fermionic theory.

Let us now comment on the connection to chaos. Since the leading pole-skipping frequency is given by $\i (l-1)2\pi T$, where $l$ is the highest spin in the theory (either integer or half-integer), for $l\ge3/2$, it seems that this frequency is positive in the imaginary direction, meaning that the Fourier mode (which is proportional to $e^{-\i\oi v}$) will grow exponentially in the retarded time. This already suggests a connection to chaos. More quantitatively, this connection was explained in \cite{Blake:2018leo} by comparing the form of the OTOC and the leading pole-skipping mode in Einstein gravity with matter. This connection was further explained in \cite{Wang:2022mcq} at the level of the metric: the shockwave solution for general higher-derivative gravity \cite{Dong:2022ucb} is found to be a limit of the quasinormal mode responsible for the leading pole-skipping point. 

In Einstein gravity or higher-derivative gravity, $l=2$ and the leading frequency is $\i \lambda_{L,\text{max}}$, where $\lambda_{L,\text{max}}=2\pi T$ is the Lyapunov exponent for maximal chaos \cite{Maldacena:2015waa}. For higher spins, the leading pole-skipping point still happens at $\i\lambda_L$ but now $\lambda_L>\lambda_{L,\text{max}}$, consistent with the well-known result that (finitely many) higher-spin fields violate the chaos bound \cite{Camanho:2014apa,Maldacena:2015waa,Maldacena:2011jn,Hartman:2015lfa}. It would be interesting to extend the connection between the leading pole-skipping point and the OTOC to $l>2$. To achieve this, one might first generalize the shockwave solution to general higher-spin theories, perhaps in the same way that shockwave solutions were constructed for a general higher-derivative theory (Appendix A of \cite{Dong:2022ucb}). The analysis in \cite{Wang:2022mcq} then suggests that the higher-spin shockwave can be obtained as a limit of the leading pole-skipping mode. Since shockwaves tell us about the OTOC, this would be enough to connect the dots.

Moreover, it would be interesting, though arguably more challenging, to do the same for half-integer $l$. In other words, can we find shockwave solutions in gravitational theories where the highest-spin field is fermionic and use that to build the connection between the leading pole-skipping frequency and the OTOC? A rigid way to compute the Lyapunov exponent in the presence of fermionic fields might require a scattering perspective \cite{Shenker:2014cwa}, and it is \emph{a priori} not clear whether there is a classical limit where the OTOC is described by a ``fermionic" shockwave. The quasinormal mode at the leading pole-skipping point, however, suggests that there might be. 

So far, this formalism has restricted to planar black holes with planar symmetry. Evidence has found that the connection between pole skipping and the OTOC holds even for rotating black holes \cite{Liu:2020yaf,Natsuume:2020snz,Blake:2021hjj,Amano:2022mlu,Jeong:2023zkf}. It would be interesting to generalize the covariant expansion formalism in this direction and use it to prove this connection for general higher-derivative gravity.

In maximally chaotic systems, the form of pole skipping and the OTOC are constrained by symmetry \cite{Blake:2017ris,Haehl:2018izb,Cotler:2018zff,Das:2019tga,Haehl:2019eae}. In the bulk, we have seen how they are constrained by the boost symmetry of the black hole. Beyond maximal chaos, the connection between pole skipping and the OTOC is more subtle \cite{Mezei:2019dfv,Choi:2020tdj,Ramirez:2020qer}. It would be interesting to understand the more general relation between them better, perhaps in effective models of non-maximal chaos \cite{Gao:2023wun,Choi:2023mab,Lin:2023trc}. 

Recent work has found pole skipping on non-black hole backgrounds \cite{Natsuume:2023hsz,Natsuume:2023lzy}. It is not clear whether there exists a similar covariant expansion formalism beyond black holes as the formalism relies heavily on the boost symmetry. However, since the example in \cite{Natsuume:2023lzy} was obtained via a double Wick rotation from a black hole spacetime, it is plausible that pole skipping happens only for those spacetimes that are related to black holes via analytical continuation, in which case the analytically continued symmetry generator might again play an important role. 

Finally, we should mention that this formalism works well in practice for any number of fields with very general interactions even though we did not present such examples. For this reason, this formalism might be helpful for interesting computations that were previously considered too complicated. For example, this could be an efficient way of constraining the quasinormal mode spectrum of a given theory \cite{Blake:2019otz}.

\acknowledgments
It is a pleasure to thank Mike Blake, Nejc \v{C}eplak, Natalia Pinzani-Fokeeva and Anthony P. Thompson for discussions on pole skipping, Xi Dong, Adolfo Holguin, Gary Horowitz and Sean McBride for discussions on spinors, and Tom Hartman, Eric Perlmutter, Douglas Stanford and Shunyu Yao for discussions on chaos and the Lyapunov exponent. SN wants to thank Joseph Conlon for his encouragement and support during the work. SN is supported by China Scholarship Council-FaZheng Group at the University of Oxford. DW is supported by NSF grant PHY2107939. ZYW is supported by the U.S. Department of Energy, Office of Science, under Award Number DE-SC0011702.

\begin{appendix}

\section{\label{app:egbose}Bosonic examples}
\subsection{Minimally coupled scalar}

Consider the following theory for a massive scalar field on the black hole background:
\begin{equation}
     \mathcal{L}=
     -\frac{1}{2}(\nabla\phi)^2-\frac{1}{2}m^2\phi^2.
\end{equation}
This has been considered in \cite{Natsuume:2019xcy} at leading weight and in \cite{Blake:2019otz} at lower weights. In this section, we perform the calculation in the covariant expansion formalism. 

Its equation of motion is given by
\begin{equation}
E\equiv \left(\nabla^2-m^2\right) \phi=0.
\end{equation}
Taking covariant derivatives in the radial direction,
\begin{align}
    (\nabla_r)^n\delta E&=(\nabla_r)^n\delta\left[\left(\nabla^2-m^2\right) \phi\right]\nn
    &=(\nabla_r)^n\left(\nabla^2-m^2\right) \delta\phi,
\end{align}
where everything is evaluated at $r=r_0$ \textit{after} derivatives are taken. Notice that we have passed the variation operator, $\di$, through the background differential operator $\left(\nabla^2-m^2\right) $ because the background is fixed. In the notation reviewed in Section~\ref{sec:bosons}, 
\begin{equation}
    \delta\mathcal{E}=[\delta E, \nabla_r\delta E,\nabla_r\nabla_r\delta E, ...]|_{r=r_0}.
\end{equation}
The components have weights $0,-1,-2,...$. In other words,
\begin{equation}
    \di \mE_0=[\di E|_{r=r_0}], \quad \di \mE_{-1}=[(\nabla_r\di E)|_{r=r_0}], \quad \text{etc.} 
\end{equation}
Since there is only one component for each weight in this example, $|\delta\mathcal{E}_p|=1$ for all $p\in \mathbb{Z}_{\le0}$. We will omit the brackets for one-by-one matrices from now on.

Since this theory has only one dynamical field which has spin zero, the highest-weight equation of motion has $p=q_0=0$:
\begin{align}
    \delta \mathcal{E}_{0}&=g^{\mu\nu}\nabla_\mu\nabla_\nu \delta \phi - m^2 \delta \phi\nn
    &=2g^{vr}\nabla_v\nabla_r \delta \phi+g^{rr}\nabla_r\nabla_r \delta \phi +g^{ij}\nabla_i\nabla_j \delta \phi- m^2 \delta \phi\nn
    &=\left(g^{ij}\nabla_i\nabla_j - m^2\right) \underbrace{\delta \phi}_{[0]}+2g^{vr}\nabla_v\underbrace{\nabla_r \delta \phi}_{[-1]}+g^{rr}\underbrace{\nabla_r\nabla_r \delta \phi}_{[-2]}\nn
    &=\begin{bmatrix}
        \left(g^{ij}\nabla_i\nabla_j - m^2\right)\quad
        & 2g^{vr}\nabla_v \quad
        & g^{rr}
    \end{bmatrix}
    \begin{bmatrix}
        \delta \phi \\
        \nabla_r \delta \phi\\
        \nabla_r^2\delta \phi
    \end{bmatrix}\nn
    &=\begin{bmatrix}
        \left(g^{ij}(\partial_i\nabla_j-\Gamma_{ij}^\m\nabla_\m) - m^2\right)\quad
        & 2g^{vr}\nabla_v \quad
        & g^{rr}
    \end{bmatrix}
    \begin{bmatrix}
        \delta \phi \\
        \nabla_r \delta \phi\\
        \nabla_r^2\delta \phi
    \end{bmatrix}.
\end{align}
From \eqref{eq:chris}, $\Gamma_{i j}^\m$ is only non-zero for $\mu=v$ (on the horizon), so
\begin{align}
    &\delta \mathcal{E}_{0}\nn
    =&\begin{bmatrix}
        \left(h^{-1}\delta^{ij}(\partial_i\partial_j+\frac{1}{2} h^{\prime} \delta_{i j}\partial_v) - m^2\right)\quad
        & 2\nabla_v \quad
        & f
    \end{bmatrix}
    \begin{bmatrix}
        \delta \phi \\
        \nabla_r \delta \phi\\
        \nabla_r^2\delta \phi
    \end{bmatrix}\nn
    =&\begin{bmatrix}
        \left(-h^{-1}k_i k_i +\frac{1}{2}h^{-1}h'(-\i\oi)d- m^2\right)\quad
        & 2(-\i\oi + \frac{1}{2}f') \quad
        & 0 
    \end{bmatrix}
    \begin{bmatrix}
        \delta \phi \\
        \nabla_r \delta \phi\\
        \nabla_r^2\delta \phi
    \end{bmatrix}.
\end{align}
Notice that we have terms with various weights $q$ on the right hand side, even though we are only considering the equation of motion with weight $p=0$. Also, in the process, we have canonicalized the expression by moving $\nabla_r$ to the right of $\nabla_v$ according to the prescription. It is easy at this order because $[\nabla_\mu,\nabla_\nu]=0$ when acting on scalars. 

Also, recall that we have defined
\begin{equation}
    \delta \mathcal{X}=
    \begin{bmatrix}
        \di\mX_{0}\\
        \di\mX_{-1}\\
        \di\mX_{-2}\\
        \cdots
    \end{bmatrix}
    =
    \begin{bmatrix}
        \di\phi\\
        \nabla_r\di\phi\\
        \nabla_r^2\di\phi\\
        \cdots
    \end{bmatrix},
\end{equation}
so what we have written down was $\di\mE_p=\sum_q\mM_{p,q}\di\mX_{q}$ for $p=0$.

Now let $\omega=-\omega_0$, then this equation reduces to
\begin{align}
    &-h^{-1}k^2-\frac{d}{2}h^{-1}h'2\pi T-m^2=0,\\
    \implies\quad &k^2 = -{d}\pi Th'-m^2h,
\end{align}
which is exactly (2.16) of \cite{Blake:2019otz}. In our language, this comes from the pole-skipping condition \eqref{psk} with $s=1$, i.e., $\det M{}_{1}(k)=0$, which in this case involves only $\mM_{0,0}$.

Let us now include the next value of $p$, which is $-1$. For simplicity, take $d=1$. Then
\begin{align}
        \delta \mathcal{E}_{-1}
        =&\,\nabla_{r}(g^{\mu\nu}\nabla_{\mu}\nabla_{\nu}-m^{2})\delta\phi\nn
        =&\,\left(\frac{2hh''-h'^2}{4h^2}\left(-2\i\omega+{f'}/{2}\right)+\frac{k^2 h'}{h^2}\right)\delta\phi\nn
        +&\,\left(\frac{f'h'}{4h}-\frac{k^2}{h}
        +\frac{h'}{2h}\left(-\i\omega+{f'}/{2}\right)+2f''-m^2\right)\nabla_r\delta\phi\nn
        +&\,2\left(-\i\omega+f'\right)\nabla_r^2\delta\phi.
\end{align}
According to \eqref{psk}, the second set of pole-skipping points appears at the frequency $\oi=-2\oi_0$ and when the determinant of $M_2(k)$ vanishes, where
\begin{equation}
    \begin{aligned}
        M_{2}(k)=\begin{bmatrix}
            -h^{-1}k^2-\frac{1}{2}{f'h^{-1}h'}-m^2&&-f'\\
            -\frac{3}{8}h^{-2}{(2hh''-h'^2)}f'+{k^2 h'h^{-2}}&&-{k^2}{h^{-1}}+2f''-m^2\\
        \end{bmatrix}.
    \end{aligned}
\end{equation}
On the BTZ background, where
\begin{align}\label{Schwarzchild3}
f(r)=r^2\(1-\frac{r_0^2}{r^2}\),~~~h(r)=r^2,
\end{align}
and for $m=0$, the pole-skipping points at this order are given by
\begin{equation}
        k^2=
            \frac{r_0^2}{2(r_0-2)} \[4\pm r_0\left(\mp 4+r_0\left(\mp 3+{r_0^{-2}}\sqrt{(4+3r_0^2)(4-8r_0+7r_0^2)}\right)\right)\].
\end{equation}
It is straightforward to continue to higher orders. We will go to higher orders in some more complicated examples.

\subsection{Maxwell theory}
\label{Maxwelltheory}
Consider Maxwell theory whose Lagrangian is given by
\begin{equation}
     \mathcal{L}=\frac{1}{4}F_{\mu\nu}F^{\mu\nu},
\end{equation}
where
\begin{equation}
    F_{\mu\nu}=\nabla_{\mu} A_{\nu}-\nabla_{\nu} A_{\mu}.
\end{equation}
This is our first example with a gauge symmetry: $A_\mu\to A_\mu+\nabla_\mu\Lambda$. This example will demonstrate how gauge redundancy is reflected in the formalism and how the gauge-covariant pole-skipping condition works.

The equation of motion is given by
\begin{equation}
\begin{aligned}
    {E}_{\nu}=\nabla^\mu F_{\mu\nu}
    =\nabla^{\mu}\nabla_{\mu}A_{\nu}-\nabla^{\mu}\nabla_{\nu} A_{\mu}.
    \end{aligned}
\end{equation}
We simplify the calculation by setting $d=1$ in this example, so
\begin{align}
      \delta {E}_v&=\nabla^{\mu}\nabla_{\mu}\delta A_v-\nabla^{\mu}\nabla_v \delta A_{\mu}\nn
      &=\nabla_v\nabla_r\delta A_v+\frac{1}{h}\nabla_x^2\delta A_v-\nabla_v^2\delta A_r-\frac{1}{h}\nabla_x\nabla_v\delta A_x.
\end{align} 
Unpacking the $\nabla$'s and evaluating it on the horizon, we have:
\begin{equation}
    \begin{aligned}
         &\delta\mathcal{E}_1=(\delta E_{v})|_{r=r_0}=(-\i\omega)\nabla_r\delta A_v-\frac{1}{h}k^2\delta A_v+\i\omega\left(-\i\omega+\frac{f'}{2}\right)\delta A_r-\frac{k\omega}{h}\delta A_x.
    \end{aligned}
\end{equation}
It is easy to see that the first pole-skipping point appears at
\begin{equation}
    (\omega,k)=(0,0)
\end{equation}
because $M_1(k)$ is just the coefficient in front of $\delta A_{v}$. We note that gauge symmetry is not yet visible at this order because $s=1$ is smaller than $q_0-u_0+1=2$, as explained in Section~\ref{sec:gauge}.

At the next order $s=2$, we have both $\nabla_r\di{E}_v$ and $\di{E}_x$ to consider. By following the prescription, we compute
\begin{align}
M_{2}(k)=\[
\begin{array}{ccc}
 -\frac{k^2}{h} & \frac{\i k f'}{2 h} & -\frac{f'}{2} \\[1.5mm]
 \frac{\i k h'}{2 h} & \frac{f' h'}{4 h} & -\i k \\[1.5mm]
 \frac{k^2 h'}{h^2} & -\frac{\i k f' h'}{2 h^2} & \frac{f' h'-4 k^2}{4 h} \\
\end{array}
\],
\end{align}
where we have chosen the basis elements in the following order:
\begin{align}
\di \mX_1\oplus\di\mX_{0}&=[\di A_v, \di A_i, \nabla_r \di A_v]|_{r=r_0}, \\
\di \mE_1\oplus\di\mE_{0}&=[\di E_v, \di E_i, \nabla_r \di E_v]|_{r=r_0}. 
\end{align}
From now on, we will only state the basis for $\di\mX$, and it should be understood that the basis for $\di\mE$ is chosen analogously, as above. The determinant of this matrix is zero for any value $k$. More precisely, this matrix has a one-dimensional kernel. This is due to the fact that the pure-gauge perturbation $\di A_\mu=\nabla_\mu \Lambda$ automatically satisfies the equations of motion. Using the language of \eqref{eq:gaugefinite}, the kernel of $M_2(k)$ is spanned by
\begin{align}
\begin{bmatrix}\di A_v \\ \di A_x\\ \nabla_r \di A_v\end{bmatrix}=\begin{bmatrix}{\i}f^\pr/2 \\ k\\ 0\end{bmatrix} \Lambda.
\end{align}
Here $\di\xi|_{r_0}=\Lambda|_{r_0}$ is the only degree of freedom of the gauge parameter at this order, so the kernel is one-dimensional. See later for cases with larger kernels. The pole-skipping points at this order can now be found using \eqref{psknew}. At
\begin{align}
k^2=-\frac{1}{4}f^\pr h^\pr,
\end{align}
the dimension of the kernel increases from $1$ to $2$.

At the next order, 
$s=3$ and $\oi=(q_0-s)\oi_0=-2\oi_0$, choosing
\begin{align}
\di \mX_1\oplus\di\mX_{0}\oplus \di\mX_{-1}&=[\di A_v, \di A_i, \nabla_r \di A_v,\di A_r, \nabla_r \di A_i, \nabla_r^2 \di A_v]|_{r=r_0}
\end{align}
as the basis, the relevant matrix is given by
\begin{align}
&M_3(k)=\\
&{\small
\[
\begin{array}{cccccc}
 -\frac{k^2}{h} & \frac{\i k f'}{h} & -f' & -\frac{1}{2} f'^2 & 0 & 0 \\[2mm]
 \frac{\i k h'}{2 h} & 0 & -\i k & \frac{1}{2}\i k f' & -f' & 0 \\[2mm]
 \frac{k^2 h'}{h^2} & -\frac{3\i k f' h'}{4 h^2} & \frac{f' h'-4 k^2}{4 h} & \frac{  f'^2 h'+2  h f' f''}{8 h} & \frac{\i k f'}{2 h} & -\frac{f'}{2} \\[2mm]
 0 & -\frac{\i k h'}{2 h^2} & -\frac{h'}{2 h} & \frac{2 h f''-f' h'-4 k^2}{4 h} & -\frac{\i k}{h} & -1 \\[2mm]
 \frac{\i k (2 h h''-3 h'^2)}{4 h^2} & \frac{h f'' h'+h f' h''-f' h'^2}{2 h^2} & \frac{\i k h'}{h} & -\frac{1}{2}\i k f'' & f''+\frac{f' h'}{2 h} & -\i k \\[2mm]
 \frac{k^2 (h h''-2 h'^2)}{h^3} & \frac{k (-h f'' h'-4 h f' h''+6 f' h'^2)}{4 \i h^3} & A & B & \frac{h k f''+2 k f' h'}{2 \i h^2} & \frac{h f''+f' h'-2 k^2}{2 h} 
\end{array}
\]}\nonumber
\end{align}
where
\begin{align}
    A&=\frac{h f'' h'+2 h f' h''-2 f' h'^2+8 k^2 h'}{4 h^2},\nn
    B&=\frac{-2 h^2 f''^2+2 h f'^2 h''-2 f'^2 h'^2-h f' f'' h'}{8 h^2},\nonumber
\end{align}
which already has a two-dimensional kernel spanned by
\begin{align}
\begin{bmatrix}\di A_v \\ \di A_x\\ \nabla_r \di A_v\\ \di A_r \\ \nabla_r \di A_x \\ \nabla_r^2 \di A_v\end{bmatrix}=\begin{bmatrix}-f^\pr \\ \i k\\ 0\\ 0 \\ -\frac{\i k h^\pr}{2h} \\ 0\end{bmatrix} \Lambda+
\begin{bmatrix}0 \\ 0\\ -\frac{1}{2}f^\pr\\ 1 \\ \i k \\ \frac{1}{2}f^\prpr\end{bmatrix} \nabla_r \Lambda,
\end{align}
where $\di\xi|_{r_0}=\Lambda|_{r_0}=\di\Xi_0$ and $\nabla_r\di\xi|_{r_0}=\nabla_r \Lambda|_{r_0}=\di\Xi_{-1}$ are the two gauge parameters appearing in the sum \eqref{eq:gaugefinite}. The pole-skipping momenta are given by the solutions to
\begin{align}
\(k^2+\frac{1}{2}f^\pr h^\pr\)(k^2+f^\pr h^\pr-h f^\prpr)-h {f^\pr}^2h^\prpr=0.
\end{align}
They will increase the dimension of the kernel from $2$ to $3$. On the BTZ black ground \eqref{Schwarzchild3}, they simplify to
\begin{equation}
    k^2= r_0^2\left(-2\pm 2\sqrt{2}\right).
\end{equation}

\subsection{Einstein gravity}
Consider the Einstein-Hilbert action with a negative cosmological constant:
\label{Einsteingravity}
\begin{align}
S=\frac{1}{16\pi G_N}\int \d^{d+2}x \sqrt{-g} (R-2\Li),~~~\Li=-\frac{d(d+1)}{2\ell^2},~~~\ell=1.
\end{align}
Compared to the other examples we study, this one is computationally the hardest. This will hopefully illustrate the advantage of using the covariant expansion method: it is fully automatable. This also serves as another illustration of the nature of gauge symmetry, which in this case is given by: $g_{\mu\nu}\to g_{\mu\nu}+\nabla_\m\zeta_\nu+\nabla_\nu\zeta_\mu$.

Einstein's equation is given by
\begin{align}
E_{\mu\nu}=R_{\mu\nu}-\frac{2\Li}{d}{g}_{\mu\nu}.
\end{align}
The linearized Einstein's equation is given by
\begin{align}
\di E_{\mu\nu}=&\,\frac{1}{2}\(-\nabla_\ai\nabla^\ai \di g_{\mu\nu}-\nabla_\mu\nabla_\nu \di g^\ai{}_\ai + \nabla_\mu \nabla^\ai \di g_{ \nu\ai}+\nabla_\nu \nabla^\ai \di g_{\mu\ai}\)\nn
+&\,\frac{1}{2}g^{\ai\bi} R_{\mu\ai}\di g_{\nu \bi}+\frac{1}{2}g^{\ai\bi} R_{\nu\ai}\di g_{\mu\bi}-g^{\ai\ri}g^{\bi\si} R_{\mu\ai\nu\bi} \di g_{\ri\si}-\frac{2\Li}{d}\di g_{\mu\nu}.
\end{align}
We will first keep $d$ general but turn to $d=2$ later for concreteness; if one wishes, one can keep $d$ general throughout the whole calculation. To avoid repetition, we now state the order in which the basis elements are presented throughout this example:
\begin{align}
[&\di g_{vv}, \di g_{v1},\di g_{v2}, \nabla_r \di g_{vv}, \di g_{vr}, \di g_{11},\di g_{12},\di g_{22}, \nabla_r \di g_{v1},\nabla_r \di g_{v2}, \nabla_r^2 \di g_{vv},\di g_{r1},\nn
&\di g_{r2}, \nabla_r \di g_{vr}, \nabla_r \di g_{11},\nabla_r \di g_{12},\nabla_r \di g_{22}, \nabla_r^2 \di g_{v1},\nabla_r^2 \di g_{v2}, \nabla_r^3 \di g_{vv},\cdots ]|_{r=r_0},
\end{align}
where the subscripts $1$ and $2$ abbreviate $x^1$ and $x^2$. To begin with, consider the highest-weight equation of motion
\begin{align}
\di\mE_2=\left.\di E_{vv}\right|_{r_0}&= \[-\frac{1}{2h}\pd_i\pd_i+\frac{d h^\pr}{4h}\(\pd_v-f^\pr\)-\frac{1}{2}f^\prpr-\frac{2\Li}{d}
 \]\di g_{vv}\nn
&~~~~~~~~~~~~~~~~+\frac{1}{h} \(\pd_v-\frac{1}{2}f^\pr\) \pd_i\di g_{vi}-\frac{1}{2h} \pd_v\(\pd_v-\frac{1}{2}f^\pr\)\di g_{ii}.
\end{align}
Again, the gauge symmetry is not visible at this order, as $s=1$ is smaller than $q_0-u_0+1=2-1+1=2$. We can therefore easily read off the location of the first skipped pole:
\begin{align}
\oi=\oi_0,~~~k^2=-\frac{d}{4}f^\pr h^\pr,
\end{align}
where we have used the fact that the background metric satisfies Einstein's equation.

At the next order ($s=2$, $\oi=0$), we have (for general $d$)
\begin{align}
\left.\di E_{vi}\right|_{r_0}&=(d-2)\frac{h^\pr}{4h}\pd_i \di g_{vv}+\[-\frac{1}{2h}\pd_j\pd_j+\frac{h^\pr}{4h}\(\pd_v-2f^\pr\)-\frac{2\Li}{d}\]\di g_{vi}+\frac{1}{2h}\pd_i\pd_j\di g_{vj}\nn
&+\frac{1}{2}\pd_i \nabla_r\di g_{vv}+\frac{1}{2}\pd_v\pd_i \di g_{vr}-\frac{1}{2}\pd_v \nabla_r \di g_{vi}+\frac{1}{2}\pd_v\(\pd_v+\frac{1}{2}f^\pr\)\di g_{ri},\\
\left.\nabla_r \di E_{vv}\right|_{r_0}&=\[-\frac{1}{2h}\pd_i\pd_i +\frac{d h^\pr}{4h} (\pd_v-f^\pr)-\frac{1}{2}f^\prpr\]^\pr \di g_{vv}-\[\frac{h^\pr}{h^2}\pd_v+\frac{1}{2}\(\frac{f^\pr}{h}\)^\pr\]\pd_i\di g_{vi}\nn
&+\[-\frac{1}{2h}\pd_i\pd_i +\frac{d h^\pr}{4h} (\pd_v-f^\pr)-\frac{1}{2}f^\prpr-\frac{2\Li}{d}\]\nabla_r\di g_{vv}+\frac{d h^\pr}{4h}\pd_v \di g_{vr}\nn
&+\frac{1}{4}\(\frac{f^\pr}{h}\)^\pr\pd_v\di g_{ii}
+\frac{1}{h}\pd_v\pd_i\nabla_r\di g_{vi}-\frac{1}{2h}\pd_v\(\pd_v+\frac{1}{2}f^\pr\)\nabla_r \di g_{ii}.
\end{align}
We can easily write down the matrix $M_{2}(k)$ using the expressions above. The determinant of this matrix is automatically zero. Its kernel is spanned by the pure-gauge perturbations $\di g_{\mu\nu}=\nabla_{\mu} \zeta_{\nu}+\nabla_{\nu} \zeta_{\mu}$ with weights $q\ge1$. In the near-horizon covariant expansion, this is
\begin{align}\label{eq:maxwellggs2}
\begin{bmatrix}\di g_{vv} \\ \di g_{vi}\\ \nabla_r \di g_{vv}\end{bmatrix}
=\begin{bmatrix}-f^\pr \\ \i k_i\\ -f^\prpr\end{bmatrix} \zeta_v,
\end{align}
where $\zeta_v|_{r_0}=\di\Xi_1$ is the only gauge parameter appearing in \eqref{eq:gaugefinite} at this order. For simplicity, we now specialize to a specific background, the Schwarzschild-AdS$_4$ black hole, which has
\begin{align}
f(r)=r^2\(1-\frac{r_0^3}{r^3}\),~~~h(r)=r^2.
\end{align}
Now our matrix simplifies to
\beq
M_2(k)=
\[
\begin{array}{cccc}
 \frac{k_1^2+k_2^2}{2 r_0^2} & -\frac{3\i k_1}{2 r_0} & -\frac{3\i k_2}{2 r_0} & 0 \\[1.5mm]
 0 & \frac{k_2^2}{2 r_0^2} & -\frac{k_1 k_2}{2 r_0^2} & \frac{\i k_1}{2} \\[1.5mm]
 0 & -\frac{k_1 k_2}{2 r_0^2} & \frac{k_1^2}{2 r_0^2} & \frac{\i k_2}{2} \\[1.5mm]
 -\frac{k_1^2+k_2^2}{r_0^3} & \frac{3\i k_1}{r_0^2} & \frac{3\i k_2}{r_0^2} & \frac{k_1^2+k_2^2}{2 r_0^2} \\
\end{array}
\].
\eeq
As we described above for general background, there is a one-dimensional kernel of this matrix \eqref{eq:maxwellggs2}. On our chosen background, it reduces to
\begin{align}
\begin{bmatrix}\di g_{vv} \\ \di g_{v1} \\ \di g_{v2}\\ \nabla_r \di g_{vv}\end{bmatrix}=\begin{bmatrix}-3 \\ \i k_1\\\i k_2\\ 0\end{bmatrix} \zeta_v.
\end{align}
The product of nonzero diagonal entries of its Jordan normal form is
\begin{align}
k^4(k^2+6r_0^2).
\end{align}
Setting $k=0$ will increase the dimension of the kernel from $1$ to $4$, while $k^2=-6 r_0^2$ will \emph{not}. According to \eqref{psknew} and with the caveat mentioned in Footnote~\ref{ft:jordan}, pole skipping only happens at $k=0$.

We can continue to find the skipped poles at the next order ($s=3$), where $\oi=-\oi_0$. We now need all equations of motion with weights greater or equal to 0. The relevant matrix is worked out to be
\begin{align}
&M_3(k)=\\
&{\scriptsize
\[
\begin{array}{cccccccccccc}
 A & -\frac{3\i k_1}{r_0} & -\frac{3\i k_2}{r_0} & 0 & 0 & -\frac{9}{4} & 0 & -\frac{9}{4} & 0 & 0 & 0 \\[1.5mm]
 0 & \frac{k_2^2}{2 r_0^2}-\frac{3}{4} & -\frac{k_1 k_2}{2 r_0^2} & \frac{\i k_1}{2} & \frac{3}{4}\i k_1 r_0 & 0 & -\frac{3\i k_2}{4 r_0} & \frac{3\i k_1}{4 r_0} & \frac{3 r_0}{4} & 0 & 0 \\[1.5mm]
 0 & -\frac{k_1 k_2}{2 r_0^2} & \frac{k_1^2}{2 r_0^2}-\frac{3}{4} & \frac{\i k_2}{2} & \frac{3}{4}\i k_2 r_0 & \frac{3\i k_2}{4 r_0} & -\frac{3\i k_1}{4 r_0} & 0 & 0 & \frac{3 r_0}{4} & 0 \\[1.5mm]
 -\frac{k_1^2+k_2^2}{r_0^3}+\frac{3}{2} & \frac{9\i k_1}{2 r_0^2} & \frac{9\i k_2}{2 r_0^2} & A & -\frac{9 r_0}{2} & \frac{9}{4 r_0} & 0 & \frac{9}{4 r_0} & -\frac{3\i k_1}{2 r_0} & -\frac{3\i k_2}{2 r_0} & 0 \\[1.5mm]
 0 & \frac{\i k_1}{2 r_0^3} & \frac{\i k_2}{2 r_0^3} & \frac{1}{r_0} & \frac{k_1^2+k_2^2}{2 r_0^2}+3 & \frac{3}{4 r_0^2} & 0 & \frac{3}{4 r_0^2} & \frac{\i k_1}{2 r_0^2} & \frac{\i k_2}{2 r_0^2} & \frac{1}{2} \\[1.5mm]
 1 & \frac{2\i k_1}{r_0} & \frac{\i k_2}{r_0} & r_0 & k_1^2+3 r_0^2 & \frac{k_2^2}{2 r_0^2}+\frac{9}{4} & -\frac{k_1 k_2}{r_0^2} & \frac{k_1^2}{2 r_0^2}+\frac{3}{4} &\i k_1 & 0 & 0 \\[1.5mm]
 0 & \frac{\i k_2}{2 r_0} & \frac{\i k_1}{2 r_0} & 0 & k_1 k_2 & 0 & \frac{3}{2} & 0 & \frac{\i k_2}{2} & \frac{\i k_1}{2} & 0 \\[1.5mm]
 1 & \frac{\i k_1}{r_0} & \frac{2\i k_2}{r_0} & r_0 & k_2^2+3 r_0^2 & \frac{k_2^2}{2 r_0^2}+\frac{3}{4} & -\frac{k_1 k_2}{r_0^2} & \frac{k_1^2}{2 r_0^2}+\frac{9}{4} & 0 &\i k_2 & 0 \\[1.5mm]
 0 & \frac{9}{4 r_0}-\frac{k_2^2}{r_0^3} & \frac{k_1 k_2}{r_0^3} & -\frac{\i k_1}{2 r_0} & \frac{3\i k_1}{4} & 0 & \frac{3\i k_2}{4 r_0^2} & -\frac{3\i k_1}{4 r_0^2} & \frac{k_2^2}{2 r_0^2}-\frac{9}{4} & -\frac{k_1 k_2}{2 r_0^2} & \frac{\i k_1}{2} \\[1.5mm]
 0 & \frac{k_1 k_2}{r_0^3} & \frac{9}{4 r_0}-\frac{k_1^2}{r_0^3} & -\frac{\i k_2}{2 r_0} & \frac{3\i k_2}{4} & -\frac{3\i k_2}{4 r_0^2} & \frac{3\i k_1}{4 r_0^2} & 0 & -\frac{k_1 k_2}{2 r_0^2} & \frac{k_1^2}{2 r_0^2}-\frac{9}{4} & \frac{\i k_2}{2} \\[1.5mm]
 \frac{3 \left(k_1^2+k_2^2-r_0^2\right)}{r_0^4} & -\frac{15\i k_1}{r_0^3} & -\frac{15\i k_2}{r_0^3} & \frac{3}{r_0}-\frac{2 \left(k_1^2+k_2^2\right)}{r_0^3} & 9 & -\frac{27}{4 r_0^2} & 0 & -\frac{27}{4 r_0^2} & \frac{6\i k_1}{r_0^2} & \frac{6\i k_2}{r_0^2} & A \nn
\end{array}
\]}
\end{align}
where
\begin{align}
    A&=\frac{1}{2} \left(\frac{k_1^2+k_2^2}{r_0^2}-3\right).\nonumber
\end{align}
Here, the dimension of the kernel is already 4, an indication of the size of the gauge group (diffeomorphism). The kernel is spanned by
\begin{align}
{\normalsize
\begin{bmatrix}\di g_{vv} \\ \di g_{v1} \\ \di g_{v2}\\ \nabla_r \di g_{vv}\\ \di g_{vr} \\ \di g_{11}\\ \di g_{12}\\ \di g_{22}\\ \nabla_r\di g_{v1}\\ \nabla_r\di g_{v2}\\ \nabla_r^2\di g_{vv}\end{bmatrix}=\begin{bmatrix}-6 \\ \i k_1\\\i k_2\\ 0\\0\\2\\0\\2\\-i k_1\\ -\i k_2\\-6\end{bmatrix} \zeta_v
+\begin{bmatrix}0 \\ -\frac{3}{2}\\0\\ 0\\0\\2\i k_1\\\i k_2\\0\\-\frac{3}{2}\\0\\ 0\end{bmatrix} \zeta_1
+\begin{bmatrix}0 \\ 0\\-\frac{3}{2}\\ 0\\0\\0\\\i k_1\\2\i k_2\\0\\-\frac{3}{2}\\ 0\end{bmatrix} \zeta_2
+\begin{bmatrix}0 \\ 0\\0\\ -3\\1\\0\\0\\0\\\i k_1\\\i k_2\\ 0\end{bmatrix} \nabla_r\zeta_v,}
\end{align}
where $\di\Xi_1=[\zeta_v]|_{r=r_0}$ and $\di\Xi_0=[\zeta_1,\zeta_2,\nabla_r\zeta_v]|_{r=r_0}$ parameterize the four dimensions of the kernel. The product of nonzero diagonal entries of its Jordan normal form is given by
\begin{align}
(k^2-3r_0^2)(k^2+6r_0^2)(k^2+9r_0^2)(k^4+9r_0^4)(k^4+15k^2r_0^2+18r_0^4).
\end{align}
Among the roots of this expression, it can be checked that $k^2=3 k_0^2$, $k^4=-9 k_0^4$ increases the dimension of the kernel.

We can continue doing this, but the size of the matrix is getting unmanageable. We will just state the results for the next two orders. At $s=4$ ($\oi=-2\oi_0$), pole-skipping momenta are given by
\begin{equation}
    k^4=18r_0^4, \quad k^4=-18 r_0^4.
\end{equation}
At $s=5$ ($\oi=-3\oi_0$), they are given by 
\begin{equation}
    k^4=27r_0^4, \quad k^4=-27 r_0^4, \quad k^2=-15r_0^2.
\end{equation}
All the results that overlap with (5.17) and (E.8) of \cite{Blake:2019otz} agree.

\section{\label{app:egfermi}Fermionic examples}

\subsection{Free Dirac spinor}

Consider a theory of a minimally coupled free Dirac field on curved spacetime:
\begin{equation}
    \mathcal{L}= \i \bar{\psi}\left(\Gamma^\m \nabla_\m-m\right) \psi.
\end{equation}
Pole skipping for this theory has been studied in \cite{Ceplak:2019ymw}. In this section, we perform the analysis using the formalism of Section~\ref{sec:fermions}. For simplicity, we work in three bulk dimensions ($d=1$). To begin with, the linear order perturbation to the equation of motion
\begin{equation}
    E =(\Gi^\mu\nabla_\mu -m)\psi
\end{equation}
is given by
\begin{align}
\di E&=\Gi^\mu\nabla_\mu \di\psi-m \di\psi\nn
&=\Gi^v \nabla_v \di\psi+\Gi^r \nabla_r \di\psi+\Gi^x \nabla_x \di\psi-m\di\psi\nn
&=\begin{bmatrix}0&\phantom{0}2\\0&\phantom{0}0\end{bmatrix}
\begin{bmatrix}(\pd_v-f^\pr/4)\di\psi_+\\(\pd_v+f^\pr/4)\di\psi_-\end{bmatrix}
+\begin{bmatrix}0&\phantom{0}f\\1&\phantom{0}0\end{bmatrix}
\begin{bmatrix}(\nabla_r\di\psi)_+\\ (\nabla_r\di\psi)_-\end{bmatrix}\nn
&+\begin{bmatrix}-1/\sqrt{h}&0\\0&1/\sqrt{h}\end{bmatrix}
\begin{bmatrix}\pd_x&-h^\pr f/(4\sqrt{h})\\ h^\pr /(4\sqrt{h})&\pd_x\end{bmatrix}
\begin{bmatrix}\di\psi_+\\\di\psi_-\end{bmatrix}-m \begin{bmatrix}\di\psi_+\\ \di\psi_-\end{bmatrix}.
\end{align}
When evaluated at the horizon,
\begin{align}
\di \mE_{1/2}=\left.\di E_+\right |_{r_0}=- \left(\frac{1}{\sqrt{h}}\pd_x +m\right)\di\psi_+ + 2\left(\pd_v+\frac{f^\pr}{4}\right)\di\psi_-.
\end{align}
According to \eqref{psk}, the first pole-skipping point happens at frequency $\oi=-\frac{1}{2}\oi_0$, and the corresponding momentum can be easily found by setting $\det M_1(k)$, which in this case is the prefactor in front of $\di \psi_+$, to zero, after substituting the Fourier expansion \eqref{eq:XFourier}. Solving for $k$ immediately leads to $k=\i m\sqrt{h}$. 

To find the next pole-skipping point, we need
\begin{equation}
    \di \mE_{-1/2} = \left.\begin{bmatrix}
        \di E_-\\
        (\nabla_r \di E)_+
    \end{bmatrix}\right |_{r_0},
\end{equation}
where
\begin{align}
\left.\di E_-\right |_{r_0}&=\frac{h^\pr }{4h}\di\psi_+ + \left(\frac{1}{\sqrt{h}}\pd_x -m\right)\di\psi_- + (\nabla_r\di\psi)_+,\\
\left.(\nabla_r \di E)_+ \right|_{r_0}&=\frac{h^\pr}{2h^{3/2}}\pd_x \di \psi_+ + \left(\frac{1}{2}f^\prpr+\frac{f^\pr h^\pr}{4h}\right)\di \psi_-
\nn
&- \left(\frac{1}{\sqrt{h}}\pd_x+m\right)\left(\nabla_r \di \psi\right)_+ +2\left(\pd_v+\frac{3}{4}f^\pr\right) (\nabla_r \di \psi)_-.
\end{align}
From the general argument, we know that setting $\omega=-\frac{3}{2}\oi_0$ would kill all terms involving $\di\mX_q$ with $q\le-3/2$. One can check explicitly that the prefactor in front of $(\nabla_r \di \psi)_-$ vanishes. As always, we are left with a square matrix to analyze:
\begin{align}
\[
\begin{array}{c}
\di E_+\\[2mm]
\di E_- \\[2mm]
(\nabla_r \di E)_+ \\
\end{array}
\]
=
\[
\begin{array}{ccc}
-\frac{1}{\sqrt{h}}\i k-m& -f^\pr & 0\\[1.5mm]
 \frac{h^\pr}{4 h} & \frac{1}{\sqrt{h}}\i k-m & 1 \\[1.5mm]
\frac{h^\pr}{2h^{3/2}}\i k & \frac{f^\prpr}{2}+\frac{f^\pr h^\pr}{4h}& -\frac{1}{\sqrt{h}}\i k-m\\
\end{array}
\]
\[
\begin{array}{c}
\di \psi_+\\[2mm]
\di \psi_- \\[2mm]
(\nabla_r \di \psi)_+ \\
\end{array}
\].
\end{align}
The determinant of the square matrix is evaluated to be
\begin{align}
\det M_{2}(k)=\frac{\i k f^\prpr}{2 \sqrt{h}}+\frac{1}{2} m f^\prpr-\frac{\i k f^\pr h^\pr}{2 h^{3/2}}-\frac{\i k^3}{h^{3/2}}-\frac{k^2 m}{h}-\frac{\i k m^2}{\sqrt{h}}-m^3.
\end{align}
On the BTZ background \eqref{Schwarzchild3}, the pole-skipping momenta can be found by solving
\begin{align}
0=\det M_2(k)=-\frac{\i}{r_0^3}(k+\i m r_0) (k-\i(m-1)r_0)(k-\i(m+1)r_0),
\end{align}
with solutions
\begin{equation}
    k=-\i mr_0,\quad \i(m-1)r_0,\quad \i(m+1)r_0.
\end{equation}

At the next order ($s=3$ and $\oi_3=-\frac{5}{2}\oi_0$), taking
\begin{align}
\di\mX_{1/2}\oplus\di\mX_{-1/2}\oplus\di\mX_{-3/2}&=[\di \psi_+, \di \psi_-, \nabla_r \di \psi_+, \nabla_r \di \psi_-, \nabla_r^2 \di \psi_+]|_{r_0}
\end{align}
as our basis (in the order presented),
\begin{multline}
    M_3(k)=\\
\left[
\begin{array}{ccccc}
 -m-\frac{\i k}{\sqrt{h}} & -2 f' & 0 & 0 & 0 \\[1mm]
 \frac{h'}{4 h} & -m+\frac{\i k}{\sqrt{h}} & 1 & 0 & 0 \\[1mm]
 \frac{\i k h'}{2 h^{3/2}} & \frac{1}{4} \left(2 f''+\frac{f' h'}{h}\right) & -m-\frac{\i k}{\sqrt{h}} & -f' & 0 \\[1mm]
 -\frac{h'^2-h h''}{4 h^2} & -\frac{\i k h'}{2 h^{3/2}} & \frac{h'}{4 h} & -m+\frac{\i k}{\sqrt{h}} & 1 \\[1mm]
 \frac{\i k \left(2 h h''-3 h'^2\right)}{4 h^{5/2}} & \frac{h \left(2 f^{(3)} h+f'' h'\right)-2 f' \left(h'^2-h h''\right)}{4 h^2} & \frac{\i k h'}{h^{3/2}} & 2 f''+\frac{f' h'}{2 h} & -m-\frac{\i k}{\sqrt{h}}
\end{array}
\right].
\end{multline}
On the BTZ background \eqref{Schwarzchild3}, the pole-skipping momenta are found by solving
\begin{align}
    0&=\det M_3(k)\\
&=\frac{-\i}{r_0^5}(k-\i mr_0)(k+\i(m-1)r_0)(k+\i(m+1)r_0)(k-i(m-2)r_0)(k-\i(m+2)r_0).\nonumber
\end{align}
At the next order ($s=4$ and $\oi_4=-\frac{7}{2}\oi_0$), taking our basis as
\begin{align}
[\di \psi_+, \di \psi_-, \nabla_r \di \psi_+, \nabla_r \di \psi_-, \nabla_r^2 \di \psi_+, \nabla_r^2 \di \psi_-, \nabla_r^3 \di \psi_+]|_{r_0},
\end{align}
the relevant matrix is given by
\begin{align}
&M_{4}(k)=\\
&{
\scriptsize
    \left[\begin{array}{ccccccc}
 -m-\frac{\i k}{\sqrt{h}} & -3 f' & 0 & 0 & 0 & 0 & 0 \\[1.5mm]
 \frac{h'}{4 h} & -m+\frac{\i k}{\sqrt{h}} & 1 & 0 & 0 & 0 & 0 \\[1.5mm]
 \frac{\i k h'}{2 h^{3/2}} & \frac{1}{4} \left(2 f''+\frac{f' h'}{h}\right) & -m-\frac{\i k}{\sqrt{h}} & -2 f' & 0 & 0 & 0 \\[1.5mm]
 -\frac{h'^2-h h''}{4 h^2} & -\frac{\i k h'}{2 h^{3/2}} & \frac{h'}{4 h} & -m+\frac{\i k}{\sqrt{h}} & 1 & 0 & 0 \\[1.5mm]
 \frac{\i k \left(2 h h''-3 h'^2\right)}{4 h^{5/2}} & A & \frac{\i k h'}{h^{3/2}} & 2 f''+\frac{f' h'}{2 h} & -m-\frac{\i k}{\sqrt{h}} & -f' & 0 \\[1.5mm]
 \frac{2 h'^3+h^2 h^{(3)}-3 h h' h''}{4 h^3} & \frac{\i k \left(3 h'^2-2 h h''\right)}{4 h^{5/2}} & -\frac{h'^2-h h''}{2 h^2} & -\frac{\i k h'}{h^{3/2}} & \frac{h'}{4 h} & -m+\frac{\i k}{\sqrt{h}} & 1 \\[1.5mm]
 \frac{\i k \left(15 h'^3+4 h^2 h^{(3)}-18 h h' h''\right)}{8 h^{7/2}} & B & \frac{3\i k \left(2 h h''-3 h'^2\right)}{4 h^{5/2}} & C & \frac{3\i k h'}{2 h^{3/2}} & \frac{3}{4} \left(6 f''+\frac{f' h'}{h}\right) & -m-\frac{\i k}{\sqrt{h}}\nonumber
\end{array}\right]
}
\end{align}
where
\begin{align}
    A&=\frac{h \left(2 f^{(3)} h+f'' h'\right)-2 f' \left(h'^2-h h''\right)}{4 h^2},\nn
    B&=\frac{3 f' \left(2 h'^3+h^2 h^{(3)}-3 h h' h''\right)+h \left(f^{(3)} h h'-3 f'' h'^2+h \left(2 h f^{(4)}(r)+3 f'' h''\right)\right)}{4 h^3},\nn
    C&=\frac{5 f^{(3)}}{2}+\frac{3 h f'' h'-6 f' \left(h'^2-h h''\right)}{4 h^2}.\nonumber
\end{align}
On the BTZ background \eqref{Schwarzchild3}, the determinant simplifies to
\begin{multline}
    \det  M_4(k) =-\frac{\i}{r_0^7}(k+\i mr_0)(k-\i(m-1)r_0)(k-\i(m+1)r_0)\\
\times (k+\i(m-2)r_0)(k+\i(m+2)r_0)(k-\i(m-3)r_0)(k-\i(m+3)r_0).
\end{multline}
Again, setting this to zero gives the corresponding pole-skipping momenta at this frequency, which one can easily read off from the expression. 

This procedure can be continued to higher orders systematically, but we will stop here to save space. To all the orders we have presented, the locations exactly match those found in \cite{Ceplak:2019ymw}.

\subsection{Rarita-Schwinger field}
\label{RSfield}

Consider the following action for the spin-$\frac{3}{2}$ Rarita-Schwinger field, $\psi_\mu$, on a curved background:
\begin{align}
S=-\frac{1}{16\pi G_N}\int \d^{d+2}x\, \sqrt{-g}\(\bar{\psi}_\mu \Gi^{\mu\nu\ri}\nabla_\nu\psi_\ri +m \bar{\psi}_\mu \Gi^{\mu\nu}\psi_\nu\).
\end{align}
This theory has been considered in \cite{Ceplak:2021efc}. This is an interesting example not only because this is the only example of ours where the dynamic field carries both Lorentz and spinor indices but also because it has a gauge symmetry only when we tune the mass $m$ to a special value.

The equation of motion for $\psi_\mu$ is given by
\begin{align}
E^\mu(\psi_\nu)=\Gi^{\mu\nu\ri}\nabla_\nu\psi_\ri+m \Gi^{\mu\nu}\psi_\nu=0.
\end{align}

This action has a gauge symmetry when $m=m_c\equiv d/2$ if the background satisfies vacuum Einstein's equation with a negative cosmological constant. To see this, consider the transformation
\begin{align}
\di \psi_\mu=\(\nabla_\mu-\frac{1}{2}\Gi_\mu \)\chi,
\end{align}
under which the equation of motion changes by
\begin{align}
\di E^\mu|_{m_c}=
&\,\Gi^{\mu\nu\ri}\nabla_\nu\(\nabla_\ri-\frac{1}{2}\Gi_\ri \)\chi+\frac{d}{2}\Gi^{\mu\nu}\(\nabla_\nu-\frac{1}{2}\Gi_\nu \)\chi\nn
=&\,g^{\mu\nu}\(R_{\nu\ri}-\frac{1}{2}g_{\nu\ri}R \) \Gi^\ri \chi-\frac{1}{2}\Gi^{\mu\nu\ri}\Gi_\ri \nabla_\nu\chi+\frac{d}{2}\Gi^{\mu\nu}\nabla_\nu \chi-\frac{d}{4}\Gi^{\mu\nu}\Gi_\nu \chi\nn
=&\,\frac{1}{2}\(R^{\mu\nu}-\frac{1}{2}g^{\mu\nu}R-\frac{d(d+1)}{2}g^{\mu\nu}\)\Gi_\nu\chi=0,
\end{align}
where $\Gi^{\mu\nu}\Gi_\nu=(d+1)\Gi^\mu$, $\Gi^{\mu\nu\ri}\Gi_\ri=d \Gi^{\mu\nu}$, and the AdS length has been set to 1. 

We again work in three bulk dimensions $(d=1)$. The highest-weight equation of motion is given by
\begin{align}
\di \mE_{3/2}=\left.\di E_{v+}\right|_{r_0}=-\left(\frac{1}{\sqrt{h}}\pd_x+m\right) \di \psi_{v+}-\frac{1}{\sqrt{h}}\left(\pd_v-\frac{1}{4}f^\pr\right)\di \psi_{x+}.
\end{align}
The first skipped pole is then located at $\oi=\frac{1}{2}\oi_0$, $k=\i m\sqrt{h}$.
To find the next few points, we need
\begin{align}\label{eq:s32M2}
\left. \di E_{v-}\right|_{r_0}&=-\frac{h^\pr}{4h}\di \psi_{v+} -\(\frac{1}{\sqrt{h}}\pd_x-m\)\di \psi_{v-}-\frac{m}{\sqrt{h}}\di \psi_{x+}+\frac{1}{\sqrt{h}}\(\pd_v+\frac{1}{4} f^\pr\)\di \psi_{x-},\nn
\left.\di E_{x+}\right|_{r_0}&=-2m\sqrt{h}\di \psi_{v-}+\sqrt{h}(\nabla_r\psi)_{v+}-\sqrt{h}\(\pd_v+\frac{1}{4}f^\pr\)\di \psi_{r+},\nn
\left. \nabla_r\di E_{v+}\right|_{r_0}&=\frac{h^\pr}{2h^{3/2}}\pd_x\di \psi_{v+}+\frac{f^\pr h^\pr}{4h}\di \psi_{v-}+\frac{f^\pr h^\pr-h f^\prpr}{4h^{3/2}}\di \psi_{x+}-\(\frac{1}{\sqrt{h}}\pd_x+m\) (\nabla_r\psi)_{v+}.
\end{align}
At the next order ($s=2$, $\oi=-\frac{1}{2}\oi_0$), in the basis
\begin{align}
\di\mX_{3/2}\oplus\di\mX_{1/2}&=[\di \psi_{v+}, \di \psi_{v-}, \di \psi_{x+}, \nabla_r \di \psi_{v+}]|_{r_0},
\end{align}
the square matrix we are interested in is given by
\begin{align}
M_2(k)=
\left[
\begin{array}{cccc}
 -m-\frac{\i k}{\sqrt{h}} & 0 & -\frac{f'}{2 \sqrt{h}} & 0 \\[2mm]
 -\frac{h'}{4 h} & m-\frac{\i k}{\sqrt{h}} & -\frac{m}{\sqrt{h}} & 0 \\[2mm]
 0 & -2 m\sqrt{h}  & 0 & \sqrt{h} \\[2mm]
 \frac{\i k h'}{2 h^{3/2}} & \frac{f' h'}{4 h} & \frac{f' h'-h f''}{4 h^{3/2}} & -m-\frac{\i k}{\sqrt{h}}
\end{array}
\right],
\end{align}
which has determinant
\begin{align}
    &\det M_2(k)\nonumber\\
    =
    &\,-\frac{k^2 f''}{4 h}-\frac{1}{4} m^2 f''+\frac{\i k m f' h'}{4 h^{3/2}}+\frac{3 m^2 f' h'}{4 h}-\frac{f'^2 h'^2}{32 h^2}+\frac{2 k^2 m^2}{h}-\frac{4 \i k m^3}{\sqrt{h}}-2 m^4.
\end{align}
As mentioned at the beginning of the section, there is a gauge symmetry for the Rarita-Schwinger field when the background Einstein's equation is satisfied. In our case, they constrain the metric as follows:
\begin{align}
\(h^\pr(r)\)^2-2h(r)h^\prpr(r)=0,~~~f^\pr(r) h^\pr(r) -4h(r)=0,~~~f^\prpr(r)=2.
\end{align}
Using the last two equations, the determinant becomes
\begin{align}
\frac{1}{2 h}(4 m^2-1)(-2\i k m \sqrt{h}-m^2 h+h+k^2).
\end{align}
For the generic case $m\ne m_c$, the determinant is non-zero and pole skipping happens at those $k$'s that make the determinant zero. For $m=m_c$, the determinant vanishes automatically, consistent with what we showed in Section~\ref{sec:gauge}. In this special case, we need to use the gauge-covariant version of pole-skipping conditions \eqref{psknew}. This gives
\begin{align}
k=-\frac{\i}{2}\sqrt{h}.
\end{align}
This momentum increases the dimension of the kernel from $1$ to $2$. We should emphasize that one cannot locate the pole-skipping points in this case by taking $m\to m_c$ after finding pole-skipping points for $m\ne m_c$, i.e., the procedures do not commute.

To understand why the matrix has a kernel of dimension 1, consider the pure-gauge perturbation
\begin{align}
\di \psi_{\mu}=\(\nabla_\mu-\frac{1}{2}\Gi_\mu \)\di\xi=\(\nabla_\mu-\frac{1}{2}\Gi_\mu \)\chi.
\end{align} 
The relevant part of \eqref{eq:ggexpn} at $\oi=-\frac{1}{2}\oi_0$ is
\begin{align}
\left[
\begin{array}{c}
\di \psi_{v+}\\
\di \psi_{v-} \\
\di \psi_{x+}\\
(\nabla_r\di \psi)_{v+}\\
\end{array}
\right]
=
-\frac{1}{2}\left[
\begin{array}{c}
 f' \\
1 \\
- (\sqrt{h}+2\i k) \\
 \frac{1}{2} f^\prpr \\
\end{array}
\right]
\chi_+.
\end{align}
This is indeed an eigenvector of the $4\times 4$ matrix \eqref{eq:s32M2} with eigenvalue 0, assuming background Einstein's equation and $m=m_c$.

At $s=3$ ($\oi=-\frac{3}{2}\oi_0$), using the basis
\begin{align}
&\di\mX_{3/2}\oplus\di\mX_{1/2}\oplus\di\mX_{-1/2}\nn
=&\,
[\di \psi_{v+}, \di \psi_{v-}, \di \psi_{x+}, \nabla_r \di \psi_{v+}, \di \psi_{x-}, \di \psi_{r+}, \nabla_r \di \psi_{v-}, \nabla_r \di \psi_{x+}, \nabla_r^2 \di \psi_{v+}]|_{r_0},
\end{align}
the matrix we are interested in is
\begin{align}
&M_3(k)=\\
&{\small
\left[
\begin{array}{ccccccccc}
 -m-\frac{\i k}{\sqrt{h}} & 0 & -\frac{f'}{\sqrt{h}} & 0 & 0 & 0 & 0 & 0 & 0 \\[1.5mm]
 -\frac{h'}{4 h} & m-\frac{\i k}{\sqrt{h}} & -\frac{m}{\sqrt{h}} & 0 & -\frac{f'}{2 \sqrt{h}} & 0 & 0 & 0 & 0 \\[1.5mm]
 0 & -2 \sqrt{h} m & 0 & \sqrt{h} & 0 & \frac{\sqrt{h} f'}{2} & 0 & 0 & 0 \\[1.5mm]
 \frac{\i k h'}{2 h^{3/2}} & \frac{f' h'}{4 h} & \frac{f' h'-h f''}{4 h^{3/2}} & -m-\frac{\i k}{\sqrt{h}} & 0 & 0 & 0 & -\frac{f'}{2 \sqrt{h}} & 0 \\[1.5mm]
 0 & 0 & 0 & 0 & 0 & \sqrt{h} m & \sqrt{h} & 0 & 0 \\[1.5mm]
 0 & 0 & -\frac{h'}{2 h^{3/2}} & 0 & \frac{2 m}{\sqrt{h}} & m+\frac{\i k}{\sqrt{h}} & 0 & -\frac{1}{\sqrt{h}} & 0 \\[1.5mm]
 \frac{h'^2-h h''}{4 h^2} & \frac{\i k h'}{2 h^{3/2}} & 0 & -\frac{h'}{4 h} & \frac{h f''+f' h'}{4 h^{3/2}} & 0 & m-\frac{\i k}{\sqrt{h}} & -\frac{m}{\sqrt{h}} & 0 \\[1.5mm]
 0 & 0 & 0 & 0 & 0 & -\frac{1}{4} \sqrt{h} f'' & -2 \sqrt{h} m & 0 & \sqrt{h} \\[1.5mm]
 \frac{\i k \left(2 h h''-3 h'^2\right)}{4 h^{5/2}} & A & B & \frac{\i k h'}{h^{3/2}} & 0 & 0 & \frac{f' h'}{2 h} & \frac{f' h'}{2 h^{3/2}} & -m-\frac{\i k}{\sqrt{h}}
\end{array}
\right]}\nonumber
\end{align}
where
\begin{align}
    A&=\frac{h f'' h'-2 f' \left(h'^2-h h''\right)}{4 h^2},\nn
    B&=\frac{h \left(f'' h'-f^{(3)} h\right)-2 f' \left(h'^2-h h''\right)}{4 h^{5/2}}.\nonumber
\end{align}
Again, the determinant is automatically zero when the background Einstein equation is satisfied and $m=m_c$. In this case, the kernel of this matrix is spanned by the pure-gauge perturbations
\begin{align}
{\small
\left[
\begin{array}{c}
\di \psi_{v+}\\
\di \psi_{v-} \\
\di \psi_{x+}\\
(\nabla_r\di \psi)_{v+}\\
\di \psi_{x-}\\
\di \psi_{r+}\\
(\nabla_r \di \psi)_{v-}\\
(\nabla_r \di \psi)_{x+}\\
(\nabla_r^2 \di \psi)_{v+}\\
\end{array}
\right]
=
\left[
\begin{array}{c}
 -f^\pr\\
 -\frac{1}{2} \\
  \frac{1}{2}\sqrt{h}+\i k \\
 -\frac{1}{4} f^\prpr \\
 \frac{h^\pr}{4 \sqrt{h}} \\
 0 \\
 0 \\
 -\frac{\i k h^\pr}{2 h} \\
 -\frac{1}{4} f^{(3)} \\
\end{array}
\right]
\chi_+
+
\left[
\begin{array}{c}
 0 \\
 -\frac{f'}{2} \\
 0 \\
 0 \\
  -\frac{1}{2} \sqrt{h}+\i k \\
 -1 \\
 \frac{f''}{4} \\
 -\frac{f' h'}{4 \sqrt{h}} \\
 0 \\
\end{array}
\right]
\chi_-
+
\left[
\begin{array}{c}
 0 \\
 0 \\
 0 \\
 -\frac{f'}{2} \\
 0 \\
 1 \\
 -\frac{1}{2} \\
 \frac{1}{2} (\sqrt{h}+2\i k) \\
 0 \\
\end{array}
\right]
(\nabla_r \chi)_+.}
\end{align}

According to our gauge-covariant pole-skipping condition \eqref{psknew}, we need to look for values of $k$ that increase the dimension of $\ker M_3(k)$. Curiously, there turns out to be none at this order. Incidentally, the values 
\begin{align}
k=\frac{\i}{2}\sqrt{h},~~~ -\frac{3\i}{2}\sqrt{h},~~~ \frac{5\i}{2}\sqrt{h}
\end{align}
would increase the number of zeros in the characteristic polynomial but not the dimension of the kernel. See Footnote~\ref{ft:jordan} for this distinction.

\end{appendix}

\bibliographystyle{JHEP}
\bibliography{bibliography}

\end{document}